\documentstyle[12pt]{article}

\newcommand{\be}{\begin{equation}}
\newcommand{\ee}{\end{equation}}
\newcommand{\bea}{\begin{eqnarray}}
\newcommand{\eea}{\end{eqnarray}}

\begin{document}
\begin{titlepage}

\begin{flushright}
{\today}
\end{flushright}
\vspace{1in}

\begin{center}
\Large
{\bf Analyticity Properties and Asymptotic  Behavior of Scattering
Amplitude  in Higher Dimensional Theories     }
\end{center}

\vspace{.2in}

\normalsize

\begin{center}
{ Jnanadeva Maharana\footnote{Adjunct Professor, NISER, Bhubaneswar}  \\
E-mail maharana$@$iopb.res.in} 
\end{center}

\normalsize

\begin{center}
 {\em Institute of Physics \\
Bhubaneswar - 751005, India  \\
    }

\end{center}

\vspace{.2in}

\baselineskip=24pt

\begin{abstract}
The properties of the high energy behavior of the scattering amplitude
 of massive, neutral and spinless particles 
in higher dimensional field theories are investigated. The axiomatic
formulation of Lehmann, Symanzik and Zimmermann is adopted. 
 The analyticity properties of the causal, the retarded and 
the advanced functions associated with
 the four point elastic amplitudes are studied.
The analog of the Lehmann-Jost-Dyson representation is obtained in  higher
 dimensional field theories. 
The generalized J-L-D representation  is utilized to derive the $t$-plane 
 analyticity property of the amplitude. 
The  existence of an ellipse analogous to the   Lehmann ellipse is 
demonstrated. 
Thus a fixed-t
dispersion relation can be written down  with  
finite number of subtractions due to the temperedness of the amplitudes.
 The domain of
analyticity of scattering amplitude in $s$ and $t$ variables is extended
 by imposing unitarity constraints.
 A generalized version of Martin's theorem
is derived to prove the existence of such a domain in D-dimensional 
field theories.
 It is shown that  
 the amplitude can be expanded in a power series in $t$ which
converges for $|t|<R$;  $R$ being
$s$-independent. The positivity properties of absorptive amplitudes are
derived  to prove the $t$-plane analyticity of amplitude. In the extended
analyticity domain dispersion relations are written with two subtractions.
 The bound on the total cross section is derived from LSZ axioms without any
extra {\it ad hoc} assumptions.

\end{abstract}

\vspace{.5in}

\end{titlepage}


{\bf 1. Introduction}

\bigskip
\noindent
The high energy scattering of hadrons is of paramount importance. The
high energy accelerators measure total cross sections, 
elastic differential cross 
sections, the characteristics of the forward diffraction peaks to
mentions a few observable quantities. On the other hand there are strong
constraints on such measurable physical quantities
 based on general frame works of
quantum field theories - the so called axiomatic field theoretic approach.
One of the celebrated achievements of the so called axiomatic field
theoretic formalism is that  the 
total cross sections in hadronic reactions 
is bounded by square of the logarithm of the
center of mass energy in the four spacetime dimensional theories. 
The experimental data for hadronic reactions respect
this bound from energy of a few GeV to highest accelerator energies.\\
Heisenberg \cite{w}, based on his deep insight and intuitions,
concluded that the total cross section for scattering of hadrons will
grow as square of logarithm of the energy. He supported his arguments in
a field theoretic model. The experimental data, at that time, limited to 
cosmic energy
data in the MeV range lent supports to his theory. 
Froissart's derivation
of the bound on total cross sections, 
that bears his name \cite{fr}, was based on
the assumption that the amplitudes satisfy the 
Mandelstam representation. However,
the Mandelstam representation has not been proved in the general frame work of
field theory. The derivation of the bound, on total cross section, which is 
based on  the rigorous results deduced from general field theory, is
credited to Martin \cite{andre}.\\  
The purpose of this investigation is to study analyticity properties of
four point scattering amplitudes of 
field theories in higher spacetime dimensions. Our intent is to derive 
rigorous results on scattering amplitude and total cross section based on
the axiomatic frame work of Lehmann, Symanzik and Zimmermann \cite{lsz}.
 The
interests in theories of higher spacetime dimensions have been increasingly
growing over a few decades.
It is believed that supersymmetry will provide an understanding of questions
related to  consistencies of the standard model such as the gauge hierarchy problem.
 There is hope
that supersymmetry might be discovered in the high energy collision experiments in 
near future. Supersymmetric theories have been  generalized to higher spacetime
dimensions subsequent to their discovery. The supergravity theories
incorporate gravity and they possess interesting attributes when generalized
to spacetime dimensions beyond four dimensions. 
Moreover, the five perturbatively consistent string theories are constructed
in ten spacetime dimensions. The string  theories hold the prospect
of unifying the fundamental forces. There are intense  research activities
and keen  interests to study
diverse aspects of theories in higher dimensions. Moreover, 
there are proposals that,
 in certain scenario,
the evidence for the existence of higher spacetime dimensions might be
revealed in the experiments of Large Hadron Collider (LHC) 
\cite{anto,luest} \footnote{see these  two articles for 
reviews and extended references}. 
In view of
such prospects it is necessary to examine what kind of precise theoretical
results can be derived in a model independent manner as far as possible.
In recent years, several important issues related to high energy gravitational
scatterings and the conceptual frame works for such processes
have drawn attentions . A 
comprehensive review of this topic will be found in Erice lectures of 
Giddings \cite{steve}.
The importance of analyticity and unitarity in higher dimensional field 
theories and their relevance in string theory has been investigated by Pius and
Sen \cite{ps}. Moreover, Irrizary-Gelpi and Siegel 
have studied the properties of
four point amplitude in higher dimensional relativistic theory in the
JWKB approximation \cite{is}. Therefore, 
there are strong motivations to study the
analyticity properties of higher dimensional field theories and derive 
rigorous results which are not based on any specific model.  Our strategy in 
deriving the analog of Froissart-Martin bound
for higher dimensional field theories is to adopt the LSZ formulation
of field theory in higher spacetime dimensions. As we shall deliberate on our
approach in sequel, we shall be guided by the developments in the 
four dimensional theories \footnote{We refer to several books, lecture notes 
and review articles here. These references provide a background on dispersion
relations, derivation of analyticity domains and Froissart bound.}
\cite{book1,book2,book3,fr1,lehm1,sommer,eden,roy} ;
 however, there are certain obstacles to be surmounted. We
shall mention them at the appropriate places.  We have adopted the axiomatic
frame work of Lehmann, Symanzik and Zimmermann (LSZ). 
It might be possible to derive
the results presented here from Wightman axioms \cite{wight,jost,streat} 
or from the more general structure
 proposed in theory  of local
observable \cite{kl,ss,bogo}. However, we adhere to LSZ formulation and endeavor to deduce
results for higher dimensional field theories. We consider scattering of 
massive, spinless, neutral particles of a single specie in D-dimensional
flat spacetime with Lorentzian signature. We also assume that there are
no bound states in this theory. 
 Our approach to the higher dimensional theories will be clear as we proceed
and we shall state our axioms in the next section.
\\
It is well known that there are host of exact results for collisions of
strongly interacting particles which are stated as theorems. These theorems are
derived under the general assumptions of quantum field theories without
appealing to any model. Notable among them is the Froissart-Martin bound
\be
\label{m-fbound}
\sigma_t(s)<{{4\pi}\over{t_0-\epsilon}}[ln(s/s_0)]^2
\ee
which restricts the growth of the total cross section, $\sigma_t$, at high
energies; $s$ being the center of mass (c.m.) energy squared and $t_0$ is
a known constant derived from field theory. 
This bound is derived from exact results which follow  from
axiomatic field theory. We shall elaborate on these aspects in Section 2. 
However, we present the essential ingredients that lead to the bound on total 
cross section:\\
 (a1) Unitarity of S-matrix.\\
(a2) The amplitude
 is analytic in
complex $cos~\theta$ inside the Lehmann-Martin ellipse;  $\theta$
being the   scattering angle in the center of
mass frame.
The focii of the ellipse lie at $(-1,+1)$ and its
semi-major axis is $cos~\theta_0 >1 $.
 The partial wave
expansion of the
amplitude, $F(s,t)$, converges absolutely inside Lehmann-Martin ellipse,
\be
\label{lexpand}
 F(s,t)={{\sqrt s}\over k}\sum _{l=0}^{\infty}(2l+1)f_l(s)P_l(cos\theta)
\ee
and  $F(s,t)$ is analytic in this region, $k$ is the c.m. momentum. 
  Unitarity  bounds  on
the partial-wave amplitudes are
\be
\label{unitaryf}
0\le |f_l(s)|^2\le Im~f_l(s)\le 1
\ee
(a3). Polynomial boundedness of the amplitude \cite{egm}.
For $0\le t< t_0$
\be
\label{polbound}
|F(s,t)|<_{|s|\rightarrow\infty}~ |s|^N
\ee
N being a positive integer.\\
(a4). $F(s,t)$ is analytic in  the complex $s$-plane. There are cuts in the
s-plane as a consequence of  $s$-channel unitarity and crossing symmetry -
crossing is  a requirement and is proved for several reactions. There are other
important bounds on the elastic differential cross sections, the width of
the diffraction peak and slope of the diffraction peak to name a few.
The statements (a1) - (a4) have been proved in the frame work of axiomatic
field theory.\\
In the context of higher dimensional field theories, the analog of Froissart
bound has been derived for scattering of massive spinless particles
\cite{moshe1,moshe2}. There is
only one scattering angle for spinless scattering although for the most
general case, the amplitude is to expressed in terms of complete set of
basis functions (see Section 4 for detailed discussions)
for representation of $SO(D-1)$ rotation group. The bound is 
\be
\label{d-bound}
\sigma_t\le C_0(lns)^{D-2}
\ee
where $C_0$ is a constant, independent of $s$. This bound was  
derived under certain reasonable assumptions
inspired by the proven results for $D=4$ field theories. Note that
for $D=4$, one recovers the high energy bound i.e.  
$\sigma_t\le {\rm Const.}(ln s)^2$. \\
Let us briefly discuss the essentials steps for derivation of (\ref{d-bound})
in \cite{moshe1, moshe2}. The  amplitude for scattering of massive spinless
particles in D-dimensional  spacetime admits  a partial wave amplitude
expansion \cite{soldate}
\be
\label{gexpand}
F^{\lambda}(s,t)=A_1s^{-\lambda+1/2}\sum_{l=0}^{\infty}(l+\lambda)f^{\lambda}_l
(s)C_l^{\lambda}(t)(1+2t/s)
\ee
where $\lambda={1\over 2}(D-3)$ and    $(s,t)$ are the usual
Mandelstam variables.
$A_1$ is a constant which is independent of $s$ and $t$; however it depends
on $\lambda={1\over 2}(D-3)$ and contains some prefactor like powers of $\pi$ and
other numerical constants. We shall display them in Section 4. Here the basis
functions are the Gegenbauer polynomials, 
$C_l^{\lambda}(x)$, and they  satisfy orthonormality properties 
 with weight factor $(1-x^2)^{{D\over 2}-2}$, $-1\le x\le+1$
\cite{szego}.
The expansion (\ref{gexpand}) converges in the domain
$-1\le cos\theta \le +1$ \cite{szego}. The prefactor $s^{-\lambda+1/2}$ is
introduced on the right hand side of (\ref{gexpand}) in order that the partial
wave amplitudes are dimensionless. Furthermore,
the partial-wave amplitudes,
 $\{ f_l^{\lambda}(s) \}$, satisfy the
unitarity constraint (\ref{unitaryf}) \cite{soldate}.
The bound (\ref{d-bound}) was derived \cite{moshe1,moshe2} with certain 
extra assumptions which had not been derived from a field theoretic basis.
\\
The two crucial
assumptions were: (AI) The amplitude
is polynomially bounded i.e.  $|F^{\lambda}(s,t)|<Cs^N$; C and N are
undetermined constants, $N$ is real positive number. (AII)
The domain of convergence of the $F^{\lambda}(s,t)$ is an extended
ellipse with  the semimajor axis
 $1+2{\tilde T}_0/s$ in the $t$-plane. The Gegenbauer polynomial is  the basis
set of functions for the case at hand and its domain of convergence
is $-1\le cos{\theta}\le +1$. Notice that ${\tilde T}_0$ is an undetermined
constant and it is independent of $s$. In other words the amplitude has a
larger analyticity domain in the  $cos{\theta}$-plane than 
$-1\le cos{\theta}\le +1$. 
Therefore, assuming the existence of the analog of the Lehmann
ellipse looks quite reasonable. Similarly, the assumption (AI) of polynomial 
boundedness for the scattering amplitude is an acceptable proposition, although
there existed no proof for it in D-dimensions. 
The authors \cite{moshe1,moshe2} derive the
above mentioned  bound (\ref{d-bound}) under assumptions (AI) and (AII). The
constant $C_0$, although is independent of $s$,  is expressed in terms of
the dimensionality of spacetime, D, the constant, ${\tilde T}_0$ and 
$N$;  thus $N$  decides
the number of subtractions required in writing the dispersion relation for the
scattering amplitude. To contrast with the Froissart-Martin bound 
(case of $D=4$), the two  
parameters ${\tilde T}_0$ and $N$ are determined from axiomatic field theoretic
consideration and the analyticity domain of the amplitude is specified. 
In most of the cases (in $D=4$), the dispersion
relations for hadronic scattering have been proved. The bound (\ref{d-bound}),
interesting as it is, lacks the rigorous basis which is known to have been
proved for 4-dimensional case. Thus it is quite desirable to make an 
endeavors to derive (AI) and (AII) from a field theoretic frame work.\\
It is pertinent to mentioned that the consequence of 
assumptions (AI) and (AII) lead to
certain important conclusions. First thing to note
 is that the partial wave amplitudes in the  expansion for
$F^{\lambda}(s,t)$ fall off exponentially beyond a cut off $L$. This is a
very crucial feature and we shall dwell upon this aspect in sequel. Moreover,
in the above mentioned work \cite{moshe1,moshe2}, as is obvious from the
bound on $\sigma_t$, the growth is still a power of $lns$.
Furthermore,  there are some interesting
results where  scattering amplitude and slope associated with the forward 
amplitude  are constrained.\\
I examined \cite{jmjmp} the scattering problem for
$D>4$ and derived some new results
in that one could obtain a upper bound for $|F^{\lambda}(s,t)|$ in a certain
domain in the complex t-plane. Furthermore, I presented a theorem
on   the distributions of zeros
of the scattering amplitude in the complex $t$-plane. Similar results
have been derived  for four dimensional theories \cite{m1}  and their
importance is well known in that context.
Moreover, with an additional assumption about the
distribution of zeros in a small domain in t-plane which includes the physical
t-region,  I derived upper and lower bounds on absorptive part of 
the scattering amplitude \cite{jmjmp}. \\
I also proposed that it might be possible to get a glimpse of the
presence of higher dimensions \cite{jmjmp} from 
the measurement of high energy total
cross section data. All theories defined in higher spacetime dimensions have
to adopt a compactification scheme in the sense that the 
radii of compactification of the extra compact dimensions are 
 very small and therefore, such length scales might not be
revealed by present high energy experiments. 
There are several compactification schemes.
Moreover, the compactification schemes
for the higher dimensional theories have revealed interesting symmetries of
the theories dimensionally reduced to lower spacetime dimensions. 
The compactification scale is
argued to be generally at higher energy scale beyond the reach of
accelerator energies in the foreseeable future. 
However, there are concrete models where the
scale of compactification is expected to be relatively low (i.e. $500$ GeV to
TeV) in the sense
that the massive  particle spectra arising from such compactification 
proposals could be observed in
Large Hadron Collider (LHC)  experiments. Therefore, it might be possible
to observe the presence of extra dimensions at LHC. I have proposed
\cite{jmjmp}  that
the energy dependence of total cross sections  might display  
such a feature that  $\sigma_t$ would 
seemingly violate the Froissart-Martin bound when very high energy
total cross section data is fitted in the  energy range
 beyond decompactification
regime such as $500$ GeV to $1$ TeV. However, the high energy bounds
in higher dimensions as alluded to above have different energy dependence
as power of $lns$.
Thus a precision analysis of data might reveal departure from the Froissart
bound and it can be interpreted as decompactification to extra dimensions.
Indeed there are specific models which advocate that the scale of
compactification could be as low as $500$ GeV to $1$ TeV and the data 
from the LHC does not
completely rule out low scale compactification models \cite{anto,luest}.
\\
If we were to pursue the proposition presented above, look for (possible)
evidence for the violation of Froissart-Martin bound in very high energy 
scatterings where extra spatial dimensions might have undergone 
decompactification then it is desirable that the constant ${\tilde T}_0$ is
determined in terms of mass scale parameters of the theory under discussion 
i.e. mass scale of decompactification or mass parameter of the theory in
question. It is obvious, this constant ${\tilde T}_0$ is not to be related
to presently determines scale (i.e. $t_0=4m_{\pi}^2$). Second issue that
deserves attention is what is the value of $N$  in (AI) which decides
the number of subtractions required to write a dispersion relation for
the amplitude (\ref{gexpand}). \\
Our attention is focused to resolve questions  alluded to in the preceding 
paragraph which are pertaining to the assumptions (AI) and (AII) stated
above. We shall work in the frame work of LSZ formulation as mentioned earlier.
Therefore, the sequence of our investigation is  close to earlier
formulations  pursued in the  the case of 4-dimensional field theory. There are 
certain difficulties when we attempt to address problems in D-dimensional 
field theory and
we shall discuss them in each of the sections as we proceed. However, we may
mention in passing 
some of these issues to illustrate the type of problems we have 
encountered. One of the important results that led Martin to extend the
domain of analyticity of scattering amplitude in $s$ and $t$ is to use
certain positivity properties of the absorptive amplitude and its
$t$-derivatives. This property was crucial to derive certain
inequalities and eventually to prove analyticity. In order to arrive at the
proof of the positivity properties, one has to resort to partial wave
expansion. We have derived these positivity relations in Section 4. 
Another important problem which deserved attention
 is to show the existence of the analog of the
Lehmann ellipses \cite{leh2}. 
We have developed  the required technical tools and
presented the derivation of the analog-Lehmann ellipse.\\
The rest of the paper is organized as follows. In the next section, we present
a short account of earlier known results of field theory which led to
derivation of Froissart-Martin bound. This section outlines the prescriptions
for derivation of the main results. We formulate the D-dimensional 
field theory following the axioms of LSZ formalism. There are no 
difficulties in defining $in$ and $out$ states for a D-dimensional theory.
 Moreover, the LSZ reduction
procedure can be adopted for the D-dimensional case to derive the expressions
for the amplitudes. We also derive expressions for the retarded function,
the advanced function and the 
causal function (which are vacuum expectation values
of distribution valued operators) and therefore the afore-mentioned 
functions are distributions. The goal is to derive the Jost-Lehmann-Dyson
representation for these distributions. We adopt a specific coordinate frame
for the D-dimensional case to obtain the Jost-Lehmann representation for
the causal function. Note that the derived J-L-D representation is expressed as
a function of Lorentz invariant variables and therefore, the results are true
in any other frame. 
Subsequently, we obtain the representation for
the retarded function. We also show how the elegant technique of Dyson can be
generalized to the D-dimensional case. The generalized Dyson's theorem and its
proof utilizing Dyson's technique is given in Appendix A in some details.
Section 3 contains the derivation of 
 the analog-Lehman ellipse for D-dimensions. This is quite
important for the derivations of generalized Martin's theorem 
in the next section. Section 4 is devoted
to study the analyticity of the amplitude for D-dimensional case. We derive the
generalized Martin's theorem in Section 4. Finally, we derive the analog
of Froissart-Martin bound in this section. It is argued that the scattering
amplitude requires at most two subtractions 
in the extended domain of holomorphy.
 We determine the semimajor axis of the large Lehmann ellipse to be
$1+{{2R}\over s}$ where $R=4m^2-\epsilon$, 
$m$ being the mass of the scalar  particle
in the D-dimensional theory. Therefore, no {\it ad hoc} parameter appear in the
expression for the bound on $\sigma_t$ in the D-dimensional theory. We 
summarise our results in Section 5. 
   In proving Martin's theorem, 
Martin had used two 'tricks' in his original paper and 
they are also presented in his 
 books \cite{book1,book2}. We have presented these as lemmas in Appendix B.  
Appendix C contains collection of some useful formulas for the Gegebauer
polynomials.
\newpage
\noindent {\bf 2. Review of Analyticity Properties of Scattering Amplitude and 
Bound on the Total Cross Section.} 
\bigskip

\noindent \noindent We review some of 
the important results of axiomatic field theory
which are necessary to derive analyticity properties of the scattering
amplitudes and to obtain asymptotic bound like the Froissart bound. Indeed,
these techniques will be implemented for higher dimensional field theories.
However, the results of four dimensional field theories are not automatically
applicable in higher dimensions as we shall discuss in the next section.
In the case of D-dimensional field theories, we encounter certain obstacles in
our intent to (eventually) derive constraints on the growth properties of the
amplitude at asymptotic energies. The resolution of these issues will be
presented at the appropriate junctures in the text. In fact the existing
results of 4-dimensional theories provide guidance for our investigations. Let
us first envisage the axioms necessary to initiate the approach in the LSZ
formulation. It is worth mentioning that the axioms stated below hold
for D-dimensional theories and these are not special features of 4-dimensional
theories.
\\
The Axioms: \\
{\bf A1.} The states of the system are represented in  a
Hilbert space, $\cal H$. All the physical observables are self-adjoint
operators in the Hilbert space.\\
{\bf A2.} The theory is invariant under inhomogeneous Lorentz transformations.\\
{\bf A3.} The energy-momentum of the states are defined. It follows from the
requirements of the Lorentz invariance that
we can construct a representation of the
orthochronous the Lorentz group. The representation
corresponds to unitary operators, $U(a,\Lambda)$,  and the theory is
invariant 
under these transformations. Thus there are hermitian operators corresponding
to spacetime translations, denoted as $P_{\mu}$ which have following
properties:
\be
\bigg[P_{\mu}, P_{\nu} \bigg]=0
\ee
If ${\cal F}(x)$ is any Heisenberg operator then its commutator with $P_{\mu}$
is
\be
\bigg[P_{\mu}, {\cal F}(x) \bigg]=i\partial_{\mu} {\cal F}(x)
\ee
The operator does not depend explicitly on spacetime 
coordinates, $x^{\mu}, \mu=0,1..D-1$.
  If one chooses a representation where the translation operators, $P_{\mu}$,
are diagonal and the basis vectors $|p,\alpha>$  span the Hilbert space,
${\cal H}$, such that
\be
P_{\mu}|p,\alpha>=p_{\mu}|p,\alpha>
\ee
then we are in a position to make more precise statements: \\
${\bullet}$ Existence of the vacuum: there is a unique invariant vacuum state
$|0>$ which has the property
\be
U(a,\Lambda)|0>=|0>
\ee
The vacuum is unique and Lorentz invariant.\\
${\bullet}$ The eigenvalue of $P_{\mu}$, $p_{\mu}$,  
is light-like, with $p_0>0$.
We are concerned  only with  massive stated in this discussion. If we implement
infinitesimal Poincare transformation on the vacuum state then
\be
P_{\mu}|0>=0,~~~ {\rm and}~~~ M_{\mu\nu}|0>=0
\ee
from above postulates. $M_{\mu\nu}$ are the generators of Lorentz
transformations.\\
{\bf A4.} The locality of theory implies that a (bosonic) local operator 
at spacetime point
$x^{\mu}$ commutes with another (bosonic) 
local operator at $x'^{\mu}$ when  their
separation is spacelike i.e. if $(x-x')^2<0$. Our Minkowski metric convention
is as follows: the inner product of two D-vectors is given by
$x.y=x^0y^0-x^1y^1-...-x^{D-1}y^{D-1}$.
Since we are dealing with a neutral scalar
field, for the field operator $\phi (x)$: ${\phi(x)}^{\dagger}=\phi(x)$ i.e.
 $\phi (x)$ is hermitian.
By definition it  transforms as a scalar under inhomogeneous Lorentz
transformations as
\be U(a,\Lambda)\phi(x)U(a,\Lambda)^{-1}=\phi(\Lambda x+a)
\ee
The micro causality can be stated as
\be
\bigg[\phi(x),\phi(x') \bigg]=0,~~~~~for~~(x-x')^2<0
\ee
It is well known that in the LSZ formalism  we are concerned with vacuum
expectation values of time ordered products of operators as well as
with the  the retarded products. The requirements of the above listed axioms
are realized
 as certain attributes of the T-products and R-products of operators.
Furthermore, the axioms also establish important relationships
   for vacuum expectation values of time ordered products and similarly
for the R-products. It is recognized that when we consider vacuum expectation
values of retarded products of field operators (the so called r-functions)
the implementation of the axioms, listed above, lead to certain linear
relations among these functions  \cite{fr1,lehm1} as we shall derive later.
 Notice that
when we impose unitarity constraints they yield nonlinear relations
among these functions. It is worth
emphasizing that separation into set of linear relations and nonlinear
relations is a hallmark of the axiomatic approach.\\
 If we contrast the LSZ
formulation with the familiar Lagrangian formalism, the (free) linear theory is
rendered trivial. Moreover, equations of motioned derived for the interacting
theory are nonlinear in the Lagrangian approach in general.
 In the case of latter,
it is not possible to state rigorously the attributes of the solutions to
the field equations. Therefore, the computation
of the S-matrix elements, in the Lagrangian formulation,  is carried out
through a chain of well defined and consistent prescriptions in the frame work
of perturbation theory \footnote{Itzykson and Zubber have discussed this
aspect in their treatise \cite{book3}}.
On the other hand in the LSZ approach the
linear relations have important consequences. \\
We proceed to state some of the salient features of the LSZ formulation. One
of the most important requirements is the asymptotic condition. This, stated
in nutshell, says that the field theory can be described in terms of
asymptotic observables which correspond to particles of definite mass and
charge. Note, however, that we are to deal with neutral massive particles.
$\phi (x)_{in}$ represents a free field and it generates a Fock space. The
dynamics is encoded in this formulation.The physical observables are
expressible in terms of the field in a unique manner. LSZ also provide a method
to relate the field $\phi_{in}(x)$ with the interacting field $\phi(x)$.
According to their formulation, $\phi_{in}(x)$ is to be defined in an
appropriate limit of $\phi(x)$. They invoke the concept of adiabatic switching
off of interaction which is another ingredient  in the LSZ approach. They
introduce the postulate of an adiabatic cut off function so that this function
controls the interactions. It is $\bf 1$ at finite time and it has smooth limit
of going to zero as $|{time}| \rightarrow \infty$. Moreover, another postulate
is that if we remove the adiabatic switching it will be possible to define
all physical quantities. The relationship  between $\phi_{in}(x)$ and $\phi(x)$
is given by
\be
\label{z}
x_0\rightarrow -\infty~~~~\phi(x)\rightarrow Z^{1/2}\phi_{in}(x)
\ee
By the first postulate, $\phi_{in}(x)$ creates free particle states. However,
in general $\phi(x)$ will create multi particle states besides the single
particle one since it is the interacting field. Moreover, $<1|\phi_{in}(x)|0>$ 
and
 $<1|\phi(x)|0>$ carry same functional dependence in $x$.  If the factor 
of $Z$ were not the scaling relation between the two
fields (\ref{z}), then canonical commutation relation for each of the 
two fields ( i.e. $\phi_{in}(x)$ and  $\phi(x)$)  will be the same.
Thus in the absence of $Z$ the two theories will be identical. Moreover, the
postulate of asymptotic condition states that in the remote future
\be
x_0\rightarrow \infty~~~~\phi(x)\rightarrow Z^{1/2}\phi_{out}(x).
\ee
Furthermore, the vacuum is unique for $\phi_{in}$,  $\phi_{out}$ and $\phi(x)$. The
normalizable single particle states are the same i.e.
$\phi_{in}|0>=\phi_{out}|0>$. We do not display $Z$ from now on. If at all
any need arises,  $Z$ can be introduced in the relevant expressions.\\
It is essential to define creation and annihilation operators for $\phi_{in}$,
$\phi_{out}$ and $\phi$. We use the plane wave basis for simplicity; however,
in a more formal approach, it is desirable to use wave packets \cite{book3}.
 Now
\be
\phi_{in}(x)={1\over{(2\pi)^{(D-1)/2}}}
\int{ d^{D-1}k\over{2|k_0|}}[e^{-ik.x}a_{in}({\bf k})
+e^{+ik.x}a_{in}^{\dagger}({\bf k}) ]
\ee
and
\be
\phi_{out}(x)={1\over{(2\pi)^{(D-1)/2}}}
\int{ d^{D-1}k\over{2|k_0|}}[e^{-ik.x}a_{out}({\bf k})
+e^{+ik.x}a_{out}^{\dagger}({\bf k}) ]
\ee
note that $\bf k$ is $(D-1)$-component spatial momentum vector of
 D-momentum,
$k$.
The operators $a_{in}({\bf k})$ and  $a_{out}({\bf k})$ and their hermitian
conjugates are postulated to be weak coupling limits of $a({\bf k}, x_0)$ and
its hermitian conjugate in the asymptotic limits,
 $x_0\rightarrow\pm\infty$ for 'in' and 'out'
operators respectively i.e.
$a_{in}({\bf k})= (weak~lim~x_0~\rightarrow -\infty)~a({\bf k},x_0)$.
Similar definition is  to be understood for creation operators for the 'in' 
case and corresponding limiting prescription is to be defined for
 'out' operators.
The mode expansion of the interacting field
$\phi(x)$ is defined below
\be
\phi(x)={1\over{(2\pi)^{(D-1)/2}}}
\int{ d^{D-1}k\over{2|k_0|}}[e^{-ik.x}a({\bf k}, x_0)
+e^{+ik.x}a^{\dagger}({\bf k},x_0) ]
\ee
It is obvious from above discussions that $\phi(x)$ interpolates between
$\phi_{in}(x)$ and $\phi_{out}(x)$ and hence the nomenclature: interpolating
field for $\phi(x)$. Moreover, as it is an interacting field, the field
equation is of the form
\be
(\Box_x-m^2)\phi(x)=j(x)
\ee
where $\Box_x$ is the D-dimensional d'Alembertian  and 
$j(x)$ is the source current operator; this is to be contrasted with
the free field equations satisfied by $\phi_{in}$ and $\phi_{out}$. We are
going to work in the Fourier (momentum) space quite often. The Fourier
transform of the current is defined as
\be
j(x)= {1\over{(2\pi)^{(D-1)/2}}}
\int d^Dke^{-ik.x}{\tilde j}(k)
\ee
The solution for $a({\bf k},x_0)$ assumes form of an integral equation
\be
a({\bf k},x_0)=a_{in}({\bf k})+\int d^Dk'\delta^{(D-1)}({\bf k}-{\bf k'})
{{e^{-i(k_0-k_0')x_0}{\tilde j}(k')}\over{k'_0-k_0+i\epsilon}}
\ee
Notice that the Fourier transformed ${\tilde j}(k)$ is well defined on the
mass shell i.e. $k^2=m^2$. We are in a position to define
 incoming and outgoing states
  using the corresponding creation operators.
\be
|k_1,k_2,....k_n~in>=a_{in}^{\dagger}({\bf k}_1)a_{in}^{\dagger}({\bf k}_2)...
a_{in}^{\dagger}({\bf k}_n)|0>
\ee
\be
|k_1,k_2,....k_n~out>=a_{out}^{\dagger}({\bf k}_1)a_{out}^{\dagger}({\bf k}_2)...
a_{out}^{\dagger}({\bf k}_n)|0>
\ee
An important comment is in order here. The generic matrix element 
$<\alpha|\phi(x_1)\phi(x_2)...|\beta>$
is not an ordinary function but a distribution. Thus it is to be always
understood as smeared with a Schwarz type test function $f\in {\cal S}$. The test
function is infinitely differentiable and it goes to zero along with all its
derivatives faster than any power of its argument. We shall derive expressions
for scattering amplitudes and the absorptive parts. It is to be understood that
these are generalized functions and such matrix elements are properly defined
with smeared out test functions.
We envisage vacuum expectation values of product operators in LSZ formulation:
either the time ordered products, the so called
T-products  or the retarded products, often denoted as R-product. We shall
be mostly concerned with the R-product throughout this investigation
\bea
R~\phi(x)\phi_1(x_1)...\phi_n(x_n)=&&(-1)^n\sum_P\theta(x_0-x_{10})
\theta(x_{10}-x_{20})...\theta(x_{n-10}-x_{n0})\nonumber\\&&
[[...[\phi(x),\phi_{i_1}(x_{i_1})],\phi_{i_2}(x_{i_2})]..],\phi_{i_n}(x_{i_n})]
\eea
with $R\phi(x)=\phi(x)$. Here P stands for all permutations $(i_1,...i_n)$ of
$1,2,...n$. The R-product is hermitial for hermitial fields $\phi_i(x_i)$ and
the product is symmetric under exchange of any fields
$\phi_1(x_1)...\phi_n(x_n)$. Notice that the field $\phi(x)$ is kept where it is
located in  its position.
We list below some of the important properties for future use \cite{fr1}:\\
(i) $R~\phi(x)\phi_1(x_1)...\phi_n(x_n) \ne 0$ only if
$x_0>~{\rm max}~\{x_{10},...x_{n0} \}$.\\
(ii) An important property of the R-product is that
\be
R~\phi(x)\phi_1(x_1)...\phi_n(x_n) = 0
\ee
whenever the time component $x_0$, appearing in the argument of $\phi(x)$ whose
position is held fix, is less than time component of any of the four vectors
$(x_1,...x_n)$ appearing in the arguments of $\phi(x_1)...\phi(x_n)$.\\
(iii) We recall that
\be
\phi(x_i)\rightarrow \phi(\Lambda x_i)=U(\Lambda,0)\phi(x_i)U(\Lambda,0)^{-1}
\ee
Under Lorentz transformation $U(\Lambda,0)$. Therefore,
\be
R~\phi(\Lambda x)\phi(\Lambda x_i)...\phi_n(\Lambda x_n)=U(\Lambda,0)
R~\phi(x)\phi_1(x_1)...\phi_n(x_n)U(\Lambda,0)^{-1}
\ee
And
\be
 \phi_i(x_i)\rightarrow\phi_i(x_i+a)=e^{ia.P}\phi_i(x_i)e^{-ia.P}
\ee
 under spacetime translations. Consequently,
\be
R~\phi( x+a)\phi( x_i+a)...\phi_n(x_n+a)=
e^{ia.P}R~\phi(x)\phi_1(x_1)...\phi_n(x_n)e^{-ia.P}
\ee
We conclude, therefore, that the vacuum expectation value of the R-product
dependents only on  difference between pair of coordinates: in other words it
depends on the
following set of coordinate differences: 
$\xi_1=x_1-x,\xi_2=x_2-x_1...\xi_n=x_{n-1} -x_n$ as a consequence of
translational invariance. \\
(iv) The retarded property of R-function and the asymptotic conditions lead
 to the following relations.
\bea
R~[\phi(x)\phi_1(x_1)...\phi_n(x_n),\phi^{in}_l(y_l)]=
i\int d^Dy'_l\Delta(y_l-y'_l)(\Box_{y'}-m_l^2)
R~\phi(x)\phi_1(x_1)...\phi_n(x_n)\phi_l(y'_l)
\eea
where $\Delta(y_l-y'_l)$ admits the representation
\be
\Delta_l(y_l-y'_l)=-{{i}\over{(2\pi)^{(D-1)2}}}\int d^Dke^{-ik.(y_l-y'_l)}
\epsilon(k_0)\delta(k^2-m_l^2)
\ee
\\
Remarks: (i) We draw attention to the fact, the axioms (A1) - (A4) introduced
in the beginning are not special to four dimensional spacetime. These axioms
are true for theories living in arbitrary spacetime spacetime dimensions, D.\\
(ii) The concept of asymptotic states and interpolating field (thus $\phi_{in}$
and $\phi_{out}$) are valid in arbitrary D. Moreover, we can construct the
Fock space from these fields as described above.\\
(iii) The definition of the retarded operator, R-product, and other features
holds in D-dimensions. Moreover, the properties of the vacuum expectation
value of R-product hold good in D-dimensions and the consequences of the
spacetime and Lorentz transformations are satisfied.\\
Therefore, adopting the LSZ reduction technique to study analyticity
properties of scattering amplitude does not encounter any problem to fulfill
the requirements of the axioms and fundamental formulation of LSZ formalism.\\

\bigskip

\noindent {\bf 2.1 The Kinematics}\\

\bigskip

\noindent We describe the kinematics for two body scattering. Although
we consider scattering of neutral, scalar particles of equal mass, we shall
continue to designate the four external particles with their momenta and denote
mass by a label. We shall use equality of mass relation whenever we so desire.
We focus attention only on $2\rightarrow 2$ elastic scattering.
 We mention is passing that for scattering of scalars in D-dimensions
the four point amplitude still depends on two Mandelstam variables $s$ and $t$
(note that for the case at hand the available variables are energy involved
in scattering and scattering angle - see more discussions later).
Our goal is to describe the essential steps used in deriving the analyticity
properties of the four point elastic amplitude in 
D-dimensions and state important results. 
\\
The D-momenta of
incoming particles are $p_a$  are $p_b$ and out going particles are $p_c$ and
$p_d$. The mass shell condition is $p_i^2=m_i^2, i=a,b,c,d$. Our convention is
that all the momenta are coming in and energy momentum conservation law is
expressed as $p_a+p_b+p_c+p_d=0$. In this convention the matrix element
for $p_a+p_b\rightarrow p_c+p_d$ is $<-p_d~-p_c~out|p_a~p_b~in>$, 
i.e. D-momenta of $c$ and $d$
are denoted with negative sign 
 from now on. The Mandelstam variables
are
\be
s=(p_a+p_b)^2,~t=(p_a+p_d)^2,~u=(p_a+p_c)^2,~
s+t+u=m_a^2+m_b^2+m_c^2+m_d^2=4m^2
\ee
It is necessary to define some more 'mass' variables for subsequent discussions
in the next section. The four 'mass' variables $M_a,M_b,M_c, M_d$ are the
lowest mass two or more particle states which have the same quantum numbers
as particles $a,b,c,d$ respectively. 
For the case at hand we define $M_a=M_b=M_c=M_d=M$
since the particles carry no internal quantum numbers. We keep carrying these
indices since we keep the option open to consider (elastic) scattering of
unequal mass particles in future 
and we might have to assign additive quantum numbers
in those cases if we so desire. There are intermediate states in  two or
more particle states in various channels (i.e. $s,t$ and $u$). Therefore,
it is necessary to define the mass variables $({\cal M}_{ab}, {\cal M}_{cd})$
$({\cal M}_{ac}, {\cal M}_{bd})$ and  $({\cal M}_{ad},{\cal M}_{bc})$.
These masses correspond to two or more particle states carrying quantum
numbers of particle pair $(ab, cd)$, $(ac, bd)$ and $(ad,bc)$. Moreover
these masses could be different in a given channel since in the general case,
the threshold for $(a,b)$
need not be same for $(c,d)$ 
although quantum numbers might be the same. The same
logic holds for the other two pairs. In our case, there is only one such mass
variable and we denote it by ${\cal M}$. It also starts from the two particle
intermediate state. 
\\
We assume that there are no bound states in the theory and consequently, there
will be no anomalous thresholds. We define
\be
s_{thr}= 4m^2={\cal M}^2,~~~{\rm and}~~~u_{thr}={\cal M}^2
\ee
and they also coincide with $s_{phys}$ and $u_{phys}$ respectively and we might
use this definition interchangeably.
\\
It is very convenient to go over to the center of mass (c.m.) system for
two body scatterings. If we denote $K_1(s)$ and $K_2(s)$ as initial and final
center of momenta for $(a,b)$ and $(c,d)$ respectively then (for equal
mass cases)

\be
K_1(s)^2={1\over 4}(s-4m^2),~~K_2(s)^2={1\over 4}(s-4m^2)
\ee
when all masses are equal and $K_1^2=K_2^2=K(s)^2$. The c.m. scattering angle is
expressed as
\be
t=-2K(s)^2(1-cos\theta), ~~~-1\le cos\theta \le +1,~~~-4K^2\le t \le 0.
\ee
We shall suppress $s$ dependence of $K$ from now on unless it is necessary.
Now we proceed to discuss the general frame work which facilitates the
investigations of analyticity properties of scattering amplitudes leading to
derivation of the Froissart-Martin bound and other such rigorous bounds.
One of the classic results is the proof of dispersion relations in the forward
direction starting from LSZ formulation and to extend it to finite interval
$-T<t\le0$, $T$ positive,  for physical $t$. 
The scattering amplitude, $F(s,t)$  is boundary
value
$ F(s,t)= {lim}_{\epsilon \rightarrow 0} F(s+i\epsilon,t)$ of $s$ and it is an
analytic function in complex s-plane with a right hand cut from 
real $s=s_{thr}$ and a left
hand cut starting from  $u=u_{thr}$. Moreover, along the left hand cut
${lim}_{\epsilon\rightarrow 0}F(s-i\epsilon, t)=F_{{a{\bar d}\rightarrow b{\bar c}}}$. 
The corresponding c.m. energy squared is $u=4m^2-s-t$. Note that for this
u-channel process $t$ represents c.m. momentum transfer squared too. Moreover,
the discontinuity across the cuts are the absorptive parts of the s-channel
and u-channel amplitudes respectively. We remark that the fixed-$t$ analyticity
properties are not sufficient to derive dispersion relations. It is necessary
to know boundedness properties of these absorptive parts. If the absorptive
amplitudes are polynomially bounded then following dispersion relation
may be written down
\bea
F(s,t,u)={{s^N}\over\pi}\int_{s_{thr}}^{\infty}{{A_s(s',t)ds'}\over{s'^N(s'-s)}}
+{{u^N}\over\pi}\int_{u_{thr}}^{\infty}{{A_u(u',t)du'}\over{u'^N(u'-u)}}+
{\rm polynomials~in~s~and~u}
\eea
The dispersion relation is expected to 
hold good for fixed $t$ as stated earlier. There
are subtleties for case of $t<0$ and finite. Recall that
$t=-2K^2(1-cos\theta) $.  
If we desire t be finite and negative (in the physical
region)  then as
$s\rightarrow s_{thr}$, $K^2 \rightarrow 0$, $cos\theta$ should be negative
and much less than unity. Lehmann attempted and  resolved this issue
successfully in the frame work of LSZ formalism. He showed that
$F(s,t,u)$ is  defined for physical s (even close to threshold) outside
   the interval $-\le cos\theta\le +1$. This complex domain is known as
small Lehmann ellipse (SLE). This result of Lehmann's 
was not adequate to resolved the
problem since even if $cos\theta$ lies inside the SLE;
 as $K\rightarrow 0$,
still $t\rightarrow 0 \sim K$ as $s$ approaches the threshold  value.
Subsequently,
Lehmann proved \cite{leh2}, in the LSZ formulation, that the absorptive part
 $A_s(s,t)$ is analytic inside a larger ellipse, the large Lehmann ellipse
(LLE), whose focii coincide with those of SLE but the semimajor axis is larger
so that he resolved the problem alluded to in the context of
$s\rightarrow {s_{thr}}$. Indeed, the powerful
theorem of Jost-Lehmann-Dyson was instrumental in proving the existence
of SLE and LLE in the LSZ formulation. Therefore,  fixed-t
dispersion relations could be written down in s for the scattering amplitude.
A further progress was made when it was demonstrated that the scattering
amplitude is analytic in $s$ and in $t$.
It is worth while to mention here that the results of Lehmann although
very important could not be utilized to derive the Froissart bound as we know it
today. The
bound derived earlier was a weaker one.\\
Another important ingredient was incorporated by Martin to derive the Froissart
bound as we know of now. He recognized the power of unitarity and
used it; especially in the context of partial wave expansion of scattering
amplitude. Note that in $D=4$,  Legendre polynomials ($P_l(cos\theta)$) are
the basis function for scattering of spinless massive particles. 
One of the crucial
component in this advancement was use of positivity properties of the
absorptive amplitude. This was  proved elegantly through the partial
wave expansion of $A_s(s,t)$. Martin, through his celebrated theorem,
proved the enlargement of domain of analyticity of the scattering amplitude.
Furthermore, he concluded that the amplitude was analytic in $s$ in the cut
plane and it was also analytic in a domain in the $t$-plane denoted by $D_t$.
These advancements in identifying the domains of analyticity of
scattering amplitude (in both $s$ and $t$)   paved the way to prove
 the Froissart-Martin bound for total
cross section (\ref{m-fbound}). The importance of this bound lies in the fact
that there is no unknown constant in (\ref{m-fbound}) except one.
Notice that the prefactor in the right hand side of (\ref{m-fbound})
is fixed in terms of the known parameters of strong interactions. However,
in the logarithm squared of $s$, one has to scale $s$ with a dimensionful
quantity: $s_0$. 
 Recently, Martin and Roy \cite{mr} have argued and shown that this scale
$s_0$ can be determined from considerations of $\pi\pi$ scattering. They 
 have put forward convincing arguments
to determine $s_0$ in terms of mass of pion i.e.
${s_0}^{-1} =17\pi{\sqrt{\pi/2}}{m_{\pi}}^{-2}$ for $\pi\pi$ scattering.
The task ahead, keeping in mind the preceding discussions, is to prove
analyticity properties of scattering amplitude for scattering of massive,
neutral particles in D-dimensions. We accept the axioms (A1) to (A4) stated
in the beginning of this section. Our procedure is to follow the formalism of
LSZ. We have argued  that the amplitudes are tempered distributions in
the D-dimensional case. The next result we need to derive is the good behavior
 of the amplitude as $s\rightarrow s_{thr}, i.e. K\rightarrow 0$ for fixed
negative t. This can be achieved if there exists analog of SLE and LLE in the
LSZ formulation of the higher dimensional theory.  As we
shall see that in order to derive the existence of SLE and LLE
 it is required that equivalent generalization of Jost-Lehmann-Dyson
theorem be proved in D-dimensional field theories.
This  is not a straight forward extension of
the $D=4$ result. We recall that the absorptive part of the amplitude
appears in the dispersion relation. As we have mentioned already, the
positivity properties of the 
absorptive part play a crucial role in deriving the
analyticity of the amplitude in $s$ and $t$ variables. Moreover,
to derive positivity properties of absorptive
amplitude, one has to take the route of
partial wave unitarity and  their positivity 
relations as was utilized by Martin.
As we shall show, in D-dimensions,
the Gegenbauer polynomials are the basis functions for partial expansions.
Thus the proof of positivity needs handling of these basis functions. Moreover,
in order to derive the analog of Froissart-Martin bound for $\sigma_t$, some
more efforts will be needed. As alluded to in the introduction, the earlier
bound on $\sigma_t$, for the D-dimensional theory,
 contained unknown parameters: one them is N, that appeared
on the polynomial boundedness property of the amplitude and a second unknown
parameter is ${\tilde T}_0$ which was introduced to define the semimajor
 axis of
the ellipse
within which the partial wave amplitude (here basis is the 
Gegenbauer polynomial)
converges. Then there is a third unknown parameter which scales the $lns$.
This scale was present in the improved proof that Martin obtained (see remarks
earlier)  for the total cross section.\\
We shall systematically proceed to obtain the necessary result to
derive analyticity and asymptotic behavior of scattering amplitude.

\newpage

\noindent {\bf 3. Analyticity Properties of Scattering Amplitude }

\bigskip
\noindent We develop the necessary formalism to study analyticity properties of
scattering of massive, neutral, spinless particles in D-dimensional spacetime 
in this section.
We have adopted the LSZ formalism and we have stated all axioms and requisite
definitions in the previous section. Thus
we begin with LSZ reduction of the four point amplitude. Let us outline
the relevant steps of reduction formula for two particle in 'in' state:
$|-p_d~-p_c~in>$. If we reduce particle 'c' then
\be
|-p_d~-p_c~in>=a^{\dagger}_c(-{\bf p}_c)|-p_d>
\ee
The state $ |-p_d~-p_c~out>$ may be reduced following an analogous
prescription. Our interest lies in evaluating the difference between the
following two four-point functions by LSZ technique
\bea
&&<-p_d~-p_c~out|p_a~p_b~in>-<-p_d~-p_c~in|p_a~p_b~in> \nonumber\\&&
=lim_{x_0\rightarrow\infty}<-p_d|a_c(-{\bf p}_c,x_0)|p_a~p_b~in>
-lim_{x_0\rightarrow -\infty}<-p_d|a_c(-{\bf p}_c,x_0)|p_a~p_b~in>\nonumber\\&&
={{i}\over{(2\pi)^{(D-1)/2}}}\int d^Dx e^{-ip_c.x}(\Box_x-m_c^2)
<-p_d|\phi_c(x)|p_a~p_b~in>
\eea
We retained the label 'c' in order to identify which particle was reduced
and we have written $m_c^2$ also for the same reason. We shall continue
to follow this convention of labeling particles and their momenta which
will serve useful purpose as will be clear soon. The above equation
is a straight forward implementation of the reduction technique. We have
two possibilities for the next step of reduction: (i) either we reduce the
single particle state $<-p_d|$  or (ii) one of the particle from
$|p_a~p_b~in>$. We end up in getting vacuum expectation value of an
R-product in either case.
\bea
&&<-p_d~-p_c~out|p_a~p_b~in>-<-p_d~-p_c~in|p_a~p_b~in> \nonumber\\&&
=-{i\over{(2\pi)^{D-1}}}\int dx^Ddy^De^{-ip_c.x-ip_b.y}(\Box_x-m_c^2)(\Box_y
-m_b^2)<-p_d|R\phi_c(x)\phi_b^{\dagger}(y)|p_a> \nonumber\\&&
=-{i\over{(2\pi)}^{D-1}}\int d^Dxd^Dye^{-ip_c.x-ip_d.y}(\Box_x-m_c^2)(\Box_y
-m_d^2)\nonumber\\&&
<0|R\phi_c(x)\phi_d(y)|p_a~p_b~in>
\eea
We have written $\phi_b^{\dagger}(y)$ deliberately to keep a tag on the field
 that it arises from reduction of 'b' in the 'in' state although we have only
neutral scalar fields.  The scattering amplitude is defined with 
the convention that
\bea
\label{f-def}
&&<-p_d~-p_c~out|p_a~p_b~in>-<-p_d~-p_c~in|p_a~p_b~in> \nonumber\\&&
=2\pi\delta^D(p_a+p_b+p_c+p_d)F(p_a,p_b,p_c,p_d)
\eea
As defined earlier the currents are $(\Box-\phi_l(x))=j_l(x),~l=a,b,c,d$. A
few comments are in order here.\\
(i) There is a subtlety involved in the operation
\be
\label{ksym}
(\Box_x-m_c^2)(\Box_y-m_d^2)(R\phi_c(x)\phi_d(y))=R(j_c(x)j_d(y)
\ee
 which
will be used in sequel. When we let $(\Box_x-m_c^2)(\Box_y-m_d^2)$ pass over
$(R\phi_c(x)\phi_d(y))$ we eventually get $R(j_c(x)j_d(y)$; 
it is to be understood that in writing this equality,
in general, there will be extra terms containing $\delta$-functions and 
the derivatives of
$\delta$-functions in such operations in addition to the term
$(R\phi_c(x)\phi_d(y))$. It has been argued by Symanzik
\cite{kurt,fr1,lehm1}
 that in a local quantum field theory only finite number of
derivatives of delta functions can appear. Therefore, when we Fourier transform
an amplitude into functions of momentum space variables (in fact functions
of Lorentz invariant variables such as $s$ and $t$ in case of four point
functions), these $\delta$-function derivatives
will appear as powers of momenta.
Therefore, these will be  only finite number of terms with powers of
momenta i.e. the amplitude will be at most
polynomials in momenta \cite{kurt}.
Indeed, the N-subtracted
dispersion relation we displayed in the previous section is justified on these
grounds when we follow the LSZ formulation. In nutshell, we see that these
amplitudes are polynomially bounded.
(ii) We may use the translation properties of the fields to simplify the above
expressions. For example, consider the product of operators $A(x)B(x')$ and
use the
translation operation on the matrix element say
$M(x,x')=<\alpha|[A(x),B(x')]|\beta>$ where $\alpha ~{\rm  and }~ \beta$
designate the momenta. Now  use translation shift by 'a'.
Then $M(x,x')=e^{-i(\beta-\alpha).a}M(x+a,x'+a)$. Choose $a=-(x+x')/2$ and
$M(x,x')$ depends on $x-x'$ as expected. Thus in host of cases, we shall see
that matrix elements depend on difference of coordinates.
Therefore, we write $Rj_l(x)j_m(x')= Rj_l(z/2)j_m(-z/2)$ where indices $l,m$
stand for $a,b,c,d$. Thus, the scattering amplitude
(\ref{f-def})  expressed as \cite{fr1,lehm1}
\be
\label{As}
F(p_a,...p_d)=-\int d^Dze^{iP.z}<-p_d|
Rj_c({z\over z})j_b^{\dagger}(-{z\over 2})|p_a>
\ee
where $P={{(p_b-p_c)}\over 2}$. In deriving (\ref{As}) we have reduced $c$ and
$b$. If we reduce $c$ and $d$ the amplitude is expressed as
\be
\label{Au}
F(p_a,...p_d)=-\int d^Dze^{-iQ.z}<0|Rj_c({z\over 2})j_d(-{z\over 2})
|p_a~p_b~in>
\ee
Now $Q$ is difference of momenta $p_c$ and $p_d$ with factor 2 dividing. We
could reduce all states of the matrix element
$<-p_d~-p_c~out|p_a~p_b~in>$ and we shall get vacuum expectation of the
R-products of four corresponding currents. This reduction is not very useful
for our investigation at the moment.
 The above expressions (\ref{As}) and (\ref{Au}) are
quite useful. The two equations derived above for the amplitude, $F$,
 are special cases of a generic retarded function
\be
\label{FR}
F_R(q)=\int d^Dze^{iq.z}\theta (z_0)
<Q_f|[j_l({z\over 2}),j_m(-{z\over 2})]|Q_i>
\ee
$j_l~{\rm and}~ j_m$ are two generic currents and indices take values
$a,b,c,d$. The two states $|Q_f>$ and $|Q_i>$ carry D-dimensional momenta
$Q_f $ and $ Q_i$ respectively and these momenta are held fixed. Thus the
argument of $F_R$ does not display $Q_f $ and $ Q_i$  and we treat
them as parameters for
the discussions to follow. We define two more functions
for our later conveniences
\be
\label{FA}
F_A=-\int d^Dze^{iq.z}\theta (-z_0)
       <Q_f|[j_l({z\over 2}),j_m(-{z\over 2})]|Q_i>
\ee
and
\be
\label{FC}
F_C(q)=\int d^Dze^{iq.z}
       <Q_f|[j_l({z\over 2}),j_m(-{z\over 2})]|Q_i>
\ee
From above definitions, it follows that
\be
\label{defFC}
F_C(q)=F_R(q)-F_A(q)
\ee
Since $F_C$ is commutator of two currents, we explicitly write the
commutator in terms of products of currents $j_l({z\over 2})j_m(-{z\over 2})$.
 Let us introduce two complete set of physical
states: $\sum_n|p_n\alpha_n><p_n\alpha_n|={\bf 1}$ and
 $\sum_{n'}|p_{n'}\beta_{n'}><p_{n'}\beta_{n'}|={\bf 1}$.
Here $\{\alpha_n, \beta_{n'} \}$
stand for quantum numbers that are permitted for the intermediate states.
Now eq. (\ref{defFC}) can be expresses as
\bea
\label{commute1}
&&\int d^Dze^{iq.z}\bigg[\sum_n\bigg(\int d^Dp_n<Q_f|j_l({z\over 2)}|p_n\alpha_n
><p_n\alpha_n|j_m(-{z\over 2})|Q_i>\bigg)\nonumber\\&& -
\sum_{n'}\bigg(\int d^Dp_{n'}<Q_f|j_m(-{z\over 2})|p_{n'}\beta_{n'}>
<p_{n'}\beta_{n'}|j_l({z\over 2})|
Q_i>\bigg) \bigg]
\eea
We may use spacetime translations on the above matrix elements of each term
to bring the arguments of the currents to $z=0$. Consequently, the intermediate
states satisfy energy momentum conservation conditions requiring
$p_n={{(Q_i+Q_f)}\over 2}-q$ and  $p_{n'}={{(Q_i+Q_f)}\over 2}+q$ and therefore,\bea
\label{F1a}
&& F_C(q)=\sum_n\bigg(<Q_f|j_l(0)|p_n={{(Q_i+Q_f)}\over 2}-q,\alpha_n><
\alpha_n,p_n={{(Q_i+Q_f)}\over 2}-q|j_m(0)|Q_i>\bigg) \nonumber\\&&
-\sum_{n'}\bigg(<Q_f|j_m(0)|p_{n'}={{(Q_i+Q_f)}\over 2}+q,\alpha_{n'}><
\alpha_{n'},p_{n'}={{(Q_i+Q_f)}\over 2}+q|j_l(0)|Q_i>\bigg)
\eea
The matrix element $F_C$ vanishes, only when each term on the right hand side
of the above equation vanishes at the same time. Therefore, 
\bea
\label{F2a}
2A_s(q)=&&\sum_{n'}\bigg(<Q_f|j(0)_l|p_{n'}={{(Q_i+Q_f)}\over 2}+q,\alpha_{n'}>
\times \nonumber\\&&
<\alpha_n',p_n={{(Q_i+Q_f)}\over 2}+q|j_m(0)|Q_i>\bigg)=0
\eea
Similarly
\bea
\label{F2b}
2A_u=&&\sum_n\bigg(<Q_f|j_m(0)|p_n={{(Q_i+Q_f)}\over 2}-q,\alpha_n>\times
\nonumber\\&&
<\alpha_n,p_n={{(Q_i+Q_f)}\over 2}-q|j_l(0)|Q_i>\bigg)=0
\eea
Thus the expressions for $2A_s$ and $2A_u$ given above must vanish
simultaneously if we desire $F_C=0$.  The intermediate states inserted in
the expressions of equations (\ref{F2a}) and (\ref{F2b}) are the  
 physical states i.e. their D-momenta must lie in the forward
light cone, $V^+$. These requirements translate to
\be
\label{minmass1}
({{Q_i+Q_f}\over 2}+q)^2\ge 0,~~~({{Q_i+Q_f}\over 2})_0+q_0\ge 0
\ee
and
\be
\label{minmass2}
({{Q_i+Q_f}\over 2}-q)^2\ge 0,~~~({{Q_i+Q_f}\over 2})_0-q_0\ge 0
\ee
Thus we should have minimum mass parameters in each of the cases which
satisfy the requirements: (i) $ ({{Q_i+Q_f}\over 2}+q)^2\ge {{\cal M}_+}^2$ and
(ii) $({{Q_i+Q_f}\over 2}-q)^2\ge {{\cal M}_-}^2 $. The matrix elements for
$A_s(q)$ and $A_u(q)$ will not vanish and if the two conditions stated above,
pertinent to each of them, are fulfilled. If we define ${\tilde F}_C(z)$ to be
the Fourier transform of $F_C(q)$
\bea
\label{FFT}
{\tilde F}_C(z)={{1}\over{(2\pi)^D}}\int d^Dqe^{-iq.z}F_C(q)
               =<Q_f|[j_m({z\over 2}),j_l(-{z\over 2})]|Q_i>
\eea
It follows from axiom of micro causality that the current commutator vanishes
outside the light cone i.e. ${\tilde F}_C(z)=0$ for $z^2<0$. Thus to repeat,
$F_C(q)\ne 0$ if one of the two conditions stated in equations (\ref{minmass1})
and (\ref{minmass2}) are satisfied.
\\
We emphasize that the retardedness property of $F_R(q)$ and similar feature of
$F_A(q)$ are crucial ingredients in order to deduce analyticity properties of
scattering amplitudes. A very important observation is, when $F_C(q)$ is zero
i.e. $F_C(q)=F_R(q)-F_A(q)=0$. Thus $F_R(q)=F_A(q)$ for those values of $q$.
This information is immensely useful to identify the analytic functions $F_R$
and $F_A$ from the generalization of reflection principle of Schwarz. The study
of the causal function, $F_C(q)$ and its analyticity properties enables
construction of $F_R(q)$ or $F_A(q)$ from the fact that $F_R(q)=F_A(q)$
whenever $F_C(q)=0$ over certain values of $q$. We may represent the retarded
function as \cite{lehm1}
\be
F_R(q)={1\over{2\pi i}}\int d^Dq'\delta^{D-1}({\bf q}'-{\bf q})
{{1}\over{(q_0'-q_0)}}F_C(q'),~~~{\rm Im}~q_0>0
\ee
In fact the above relationship is more transparent if we go over to the
coordinate space through a Fourier transform  and note
\be
\label{edge}
{\tilde F}_R(z)=\int d^Dqe^{-iq.z}F_R(q)=\theta(z_0)<Q_f|[j_l({z\over 2}),
j_m(-{z\over 2})|Q_i>=\theta(z_0){\tilde F}_C(z)
\ee
Let us consider a specific case where we identify $|Q_i>=|p_a>$ and
$|Q_f>=|-p_d>$. Therefore, we have reduced 'b' and 'c' and the associated
currents in the R-product matrix elements are:
$j_l({z\over 2})=j_c({z\over 2})$ and
$j_m(-{z\over 2})=j_b^{\dagger}(-{z\over 2})$ (we continue to write
$j^{\dagger}$). A few remarks are called for at this stage.\\
(i) We have noted how the matrix element ${\tilde F}_C$ vanishes outside the
light cone.\\
(ii) We observed that for certain values of $q$, $F_C(q)$ vanishes and
consequently, $F_R(q)$ and $F_A(q)$ coincide there. We recall that in the
context of $D=4$, the edge-of-the-wedge theorem plays a powerful role
in the study of the four point function (with four momenta ($p_a,p_b,p_c,p_d$)).
The amplitude is uniquely represented by analytic function of these
complexified momenta. The amplitude is an analytic function on the manifold
$p_a+p_b+p_c+p_d =0$. The method to find the domain of holomorphy is termed
as linear problem since unitarity condition is not invoked. Bremermann, Oehme
and Taylor \cite{bmt}
proved the edge-of-the-wedge theorem for 4-point function in the
LSZ frame work.\\
(iii) We have not furnished detail proof of edge-of-the-wedge theorem
 for the massive
scalar theory in D-dimensions. However, it is quite conceivable that the proof
of Bremermann, Oehme and Taylor \cite{bmt}
is likely to go through. It seems there are no 
 serious obstacles in generalizing the theorem to D-dimensions.
Let us recapitulate the essential arguments of Bremermann, Oehme and
Taylor \cite{bmt}.
The Appendix of their  paper proves Lemma 1 and Lemma 2 prior to proving the
edge-of-the-wedge theorem. 
We briefly outline the content of Lemma 1 of \cite{bmt}; we refer the reader to
the appendix of the paper for details.  $f(z_0,z_1)$ is a 
function of two complex variables $(z_0,z_1)$ and is given as a Fourier
transform of two tempered distributions. The function is analytic in the
"wedge" W defined below
\be
\label{wedge1}
W=\bigg[(z_0,z_1): |y_1|<|y_0|,~|x_0|<\infty,~|x_1|<\infty \bigg]
\ee
where $(x_0,x_1)$ and $(y_0,y_1)$ are real and imaginary parts of $(z_0,z_1)$
respectively. Let $E$ be a given domain in $(x_0,x_1)$ plane. The authors
define the "E-limiting sequence" for a pair of complex numbers 
$(z_{0n},z_{1n})$ if they satisfy the following conditions: (a) 
${lim_n}~y_{0n}={lim_n}~y_{1n}=0$. (b) $lim_n~(x_{0n},x_{1n})\in E$.
(c) There is a number $c>1$, independent of $n$, so that for all $n$
$|y_{0n}|>|y_{1n}|$. It is then assumed that$f(z_0,z_1)$ has the limiting
property that for any E-limiting sequence; the limit $lim_n~f(z_{0n},z_{1n})$
exists. It is independent of the particular sequence and depends on the
limit point. Then BOT \cite{bmt} proved that if $f(z_0,z_1)$ is analytic
in some neighborhood $N$ of the set
$S=[(z_0,z_1):y_0=y_1=0,~(x_0,x_1)\in E]$. There was a choice of a coordinate
system such that $(x_0,x_1)=(0,0)$ to be  a particular point of $E$ and 
analyticity was proven at $z_0=z_1=0$. The authors assumed that
$f(z_0,z_1)$ is analytic in the neighborhood of  $z_0=z_1=0$ and therefore,
the power series expansion 
(like $f(z_0,z_1)=\sum_{m=n=0}^{\infty}a_{mn}z_0^mz_1^n$) exists for some $r$,
$|z_0|<r$, $|z_1|<r$ and the expansion uniformly convergent.  One can
define an analytic plane 
${\bf \pi}_{\alpha}: z_0=\alpha_0\lambda,~z_1=\alpha_1\lambda$; 
$(\alpha_0,\alpha_1)$ real and $|{{\alpha_0}\over{\alpha_1}}|<1$. Eventually,
BOT proved that power series expansions on analytic planes ${\bf\pi}_{\alpha}$
(these planes can be suitably defined) can be joined together to give a
power series expansion (this is defined in specific domains for 
$(z_0$ and $z_1)$)  which equals the separate power series expansions and 
therefore, is equal to  $f(z_0,z_1)$ from where one started with. Thereby,
the analyticity of  $f(z_0,z_1)$ is proved in the neighborhood of 
$z_0=z_1=0$. This Lemma is subsequently used to prove edge-of-the-wedge 
theorem for the  four point amplitude of equal mass scattering  of scalars
in four dimensions. 
 In the Lemma 2,  they
consider a function $f(z)$ which is function of four complex variables i.e.
$(z=z_0,z_1,z_2,z_3= z_0, {\bf z})$. Thus by definition $\bf z$ has three
components. Now define a wedge 
 $W$ for the system of  four complex variables.
\be
\label{wedge2}
W=\bigg[(z_0,{\bf z}):|y_0|>|{\bf y}|,~|x_0|<\infty,~|{\bf x}|<\infty \bigg]
\ee
 Next they extend the proof of analyticity of properties $f(z)$ as
a function of four complex variables along the technique utilized in the
proof of Lemma1. In order to prove the edge-of-the-wedge theorem for the
4-dimensional case, they define a function $f(z,z')$ as a function of $8$
complex variables, $z=(z_0,z_1,z_2,z_3)= (z_0, {\bf z})$ and
 $z'=(z'_0,z'_1,z'_2,z'_3)= (z'_0, {\bf z}')$  and  define two wedges
associated with each of the four complex variables. 
It is assumed  that  $f(z,z')$ is
analytic in the double wedge $W\otimes W$. Furthermore, 
this function is Fourier
transform of tempered distributions. Subsequently, these authors \cite{bmt}
prove that
this function (of eight complex variables) can be analytically continued and
they arrive at the proof of the edge-of-the-wedge theorem. In four dimensions,
there are two momentum variable for the problem (each momentum has four
components) and when complexified it gives rise 
 to eight complex variables. Therefore, intuitively, it looks plausible
that the proof might go through when we deal with amplitude in D-dimensional
field theory. 
 Therefore, existence of the proof can be taken as a well
judged conclusion.\\
(iv) Bros, Epstein and Glaser \cite{beg1} studied analyticity domain of
the four point amplitude in complex four momentum space. They adopted a
geometrical technique for analytic completion\footnote{ see Martin
\cite{book2} for lucid exposition to technique of analytic completion and
for illustrative examples.}. They derived analyticity in
both the variables $s$ and $t$ for the on mass 
shell amplitude. As noted earlier, we deal with the Fourier transforms of
the vacuum expectation values corresponding
to retarded and advanced products of field operators. The Fourier transformed
functions are analytically continued (the functions in the $x$-space are
tempered distributions). Furthermore, the functions defined in the 
momentum space coincide in the coincidence region. It follows from the
edge-of-the-wedge theorem that amplitudes for various  of four particle
reactions are represented by by boundary values of unique analytic functions
depending on three independent complex variables. 
The $2\rightarrow 2$ amplitudes 
are functions ${\cal F_I}(p_a,p_b,p_c,p_d)$ with $\sum p_l=0, l=a,b,c,d$;
therefore, the amplitude is effectively a function of three complexified
variables.  
There are many more functions besides $F_R$, $F_A$ and $F_c$. We have
discussed the support properties of these three functions already.
${\cal F_I}(p_a,p_b,p_c,p_d)$ is boundary value of a function
${\cal F_I}(k_a,k_b,k_c,k_d)$, $k_l=p_l+iq_l, l=a,b,c,d$. 
 ${\cal F_I}(k_a,...)$ is analytic in the tube 
${\cal T}_A=\{ k, q\in V^+_A\}$; such that $\sum k_l=0$. Here  $V^+_A$
is the forward light cone. Moreover, due to the
spectrum conditions, these functions
${\cal F_I}(k_a,k_b,k_c,k_d)$ coincide in some real regions of the momentum
space (${\rm Im}~k_l=0,~ l=a,b,c,d$). Therefore, by the
edge-of-the-wedge theorem, there exists a function $H(k_a,k_b,k_c,k_d)$ 
which is their analytic continuations. It is natural to ask whether the
formalism of \cite{beg1} is specifically applicable to four dimensional
theories if we restrict the analysis to four point amplitude.
We argue that  
the BEG \cite{beg1} procedure may  be applied to the four point amplitude
in D-dimensional field theories.
 In their more general formulation, BEG  were also 
 working within the LSZ frame work. It is not very clear if the crucial
fact that they were working in $D=4$ is essential for their proof.
In the case of D-dimensional spacetime, the BEG theorem \cite{beg1} might be
proved for the four point amplitude. 
In the case of four point function, let us choose a coordinate system 
where particles $a,b,c,d$ are assigned following momenta:
\bea
\label{begx}
&& p_a=(p_{a0},p_{a1},p_{a2},p_{a3}, {\bf 0}) \nonumber\\&&
p_b=(p_{b0},p_{b1},p_{b2},p_{b3}, {\bf 0}) \nonumber\\&& 
p_c=(p_{c0},p_{c1},p_{c2},p_{c3}, {\bf 0}) \nonumber\\&&
p_d=(p_{d0},p_{d1},p_{d2},p_{b3}, {\bf 0}) 
\eea
where $\bf 0$ stands for  the $D-3$ dimensional spatial components of the
$D-1$ spatial vectors. We can always choose the $D$-momentum vectors of the
four particles in this manner.  The BEG \cite{beg1} proof of the 
edge-of-the-wedge theorem will go through as long as we consider the four
point amplitude.\footnote{ I thank H. Epstein for very illuminating discussions
on this point and in advancing these arguments.}
 However, in $D-dimensions$ the arguments of BEG cannot be
extended for an arbitrary $n$-point amplitudes in this manner in general.
 Moreover, the four point amplitude depends
only on the Lorentz invariant variables $s$ and $t$. Thus if we obtain
the BEG proof of the edge-of-the-wedge theorem in this special choice
of (momentum) frame, then it will be 
valid in any Lorentz frame. However, when BEG
argument is adopted in this frame, we are to implement the analytic 
continuation in this four dimensional subspace from one domain to another
domain following the arguments used for $D=4$ theories \cite{beg1,ep}. 
Let us consider the simple case of n-point amplitude where $4<n<D$. Here 
we are unable to restrict the D-momenta of n-particles to a four dimensional
subspace to apply BEG theorem (proved for $D=4$ theories).
 We cannot  offer arguments for the
proof of the edge-of-the-wedge theorem for higher point amplitudes in 
$D$-dimensions as we have advanced for the four point amplitude.  \\ 
(v) It is worth while to point out the approach of
Bogoliubov and collaborators \cite{bogo,bogo1}. In this formalism, 
the proof of 
non-forward dispersion relation does not make use of the theory of several
complex variables as was adopted by \cite{bmt}. The formalism   of this group 
\cite{bogo,bogo1} is to employ certain parametrizations and the technique of 
distributions.  Their formalism is mathematically rigorous.
In their treatise on axiomatic field theory \cite{bogo}, they present the formalism
to prove edge-of-the-wedge theorem. They develop a theory for study of
the analyticity properties of a function of $n$-complex variable
systematically. Then prove analog of the edge-of-the-wedge theorem before
resorting to usual $4$-dimensional spacetime case. 
It is quite possible that their
technique might be useful to prove 
the edge-of-the-wedge theorem in D-dimensions.\\
(vi) Let us focus attention on $F_R(q)$. It is a function of D-component
complex vector q. We remind that, in coordinate space expression, it vanishes
outside the light cone. Moreover, the function is defined in the future light
cone: $V^+=\{z~|z^2>0, ~z_0>0 \}$. The matrix element is tempered. The
expression will converge (thus $F_R(q)$ will be analytic in $q$) if in the
Fourier transform $e^{iq.z}$ falls off exponentially for $z\rightarrow\infty$
in all directions such that $z$ is in the support of the integrand. It is
evident that we must have $z.{\rm Im}~q>0$ and at the same time all $z\in V^+$.
Moreover, we should have $Im~q\in V^+$. Now define
\be
\label{Tube}
T^+=\{q~|~(Im~q)^2>0,~~Im~q_0>0,~~{\rm Re}~q~{\rm arbitrary}\}
\ee
This is the definition of  the forward tube. Thus $F_R(q)$ is holomorphic in $q$ for
$q\in T^+$. We may go through the same arguments and conclude that $F_A(q)$ is
also holomorphic in $q$ for $q\in T^- = -T^+$. The region where $F_R(q)=F_A(q)$
i.e. where $F_C(q)=0$ corresponds to the domain where $q$ is real ($Im~q=0$).
If we invoke the preceding arguments, we may conclude that $F_R(q)$ and
$F_A(q)$ are the same analyitic functions. This will follow from the
edge-of-the-wedge theorem generalized to D-dimensions.
In fact this identification is intimately related to reflection principle of
Schwarz.\\
We have stated earlier that there are no bound states in the theory and
$s_{thr}=s_{phys}$; furthermore, $u_{thr}=u_{phys}$. For elastic scattering,
the right hand cut starts at $s=s_{thr}=4m^2$ and the left hand cut at
$u=u_{thr}$. In order to study further the analyticity property of the
amplitude, it is necessary to obtain a representation for $F_C(q)$ and then
we can also derive representation for $F_R(q)$. The constraint of
micro causality on ${\tilde F}_C(z)$ plays an important role in this
process.

\bigskip

\noindent {\bf 3.2 The Jost-Lehmann-Dyson Representation}

\bigskip
\noindent Jost and Lehmann \cite{jl}
obtained an integral presentation for the matrix
element of the causal commutator. This is a very powerful result. The
analyticity properties of the scattering amplitude, as a function of $s$ and
$t$, can be investigated. These are of interests to us and we shall generalize
the known result to the D-dimensional case. Dyson \cite{dyson}
introduced a very powerful
 technique and derived  more general results in that the Jost-Lehmann
representation is valid for the case of equal mass particles and had some
limitations. On the other hand Dyson's method is applicable for the case of
unequal masses in a more general setting.
 We have generalized Dyson's theorem for the case
of D-dimensional spacetime and derived the corresponding representation for
$F_C(q)$ and $F_R(q)$
and the results are presented in the Appendix A. \\
We have   stated earlier the properties satisfied by ${\tilde F}_C(x)$ (see
eq. (\ref{FFT}) and the discussions that follow). The region where $F_C(q)$
is nonvanishing is specified through equations (\ref{minmass1}) and
(\ref{minmass2}). Let us designate this domain as ${\bf{\bar R}}$
\bea
\label{def-R}
{\bf{\bar R}}:\bigg\{(Q+q)^2 \ge{\cal M}_+^2,~Q+q\in V^+~~{\rm and}~~
   (Q-q)^2 \ge{\cal M}_-^2,~Q-q\in V^+ \bigg\}
\eea
as stated earlier $V^+$ is the future light cone. The Jost-Lehmann
representation for $F_C(q)$ is such that it is nonvanishing  in the region
${\bf\bar R}$ given by (\ref{def-R}) and the Fourier transform vanishes outside
 light cone
\be
\label{jostleh}
F_C(q)=\int_Sd^Du\int_0^{\infty}d\chi^2\epsilon(q_0-u_0)\delta[(q-u)^2-\chi^2)]
\Phi(u,Q.\chi^2)
\ee
Note that $u$ is also a D-dimensional vector ({\it no relations with Mandelstam
variable u}). The domain of integration of $u$ is the region $S$ specified
below
\bea
\label{domnS}
{\bf S}:\bigg\{Q+u \in V^+,~ Q-u \in V^+,~
Max~ [0,{\cal M}_+-\sqrt{(Q+u)^2},{\cal M}_--\sqrt{(Q-u)^2}]\le \chi \bigg\}
\eea
and $ \Phi(u,Q.\chi^2)$ arbitrary.
 Here $\chi^2$ is like a mass parameter. Notice that
the assumptions about the features of the causal function stated above are
the properties we have listed earlier if we identify $Q={{(Q_i+Q_f)}\over 2}$.
In order to obtain a representation for the retarded function, we recall
that $F_R(q)$ and $F_C(q)$ are related by \cite{jl}
\be
\label{rjostleh}
F_R(q)={{i\over {2\pi}}}\int d^Dq'\delta^{D-1}({\bf q'}-{\bf q})
{{1\over{q_0'-q_0}}}F_C(q'), ~Im~q_0>0
\ee
Therefore, the Jost-Lehmann representation for $F_R(q)$ reads \cite{jl}
\be
\label{jostlehFR}
F_R(q)={{i\over{2\pi}}}\int_Sd^Du\int_0^{\infty}d\chi^2
{{\Phi(u,Q,\chi^2)}\over{(q-u)^2-\chi^2}}
\ee
This integral representation is valid provided the integral converges. We
have noted earlier in derivation of the expression for ${\tilde F}_R(z)$
that it is defined with the understanding that there could be additional
terms corresponding to $\delta$-functions and their finite number of
derivatives. The support for this argument is that in a local field theory
only finite number of such derivatives could occur \cite{kurt}.
 Therefore, the above
integrand (it is defined now in the momentum space) might have at most
polynomials in momentum which can be taken care of by appealing to the
subtraction prescription. However, the analyticity properties are unaffected
by subtractions.\\
An important point to note that the singularities are in the complex plane
as is obvious from (\ref{jostlehFR}). These points are solution to
 the equation
\be
\label{FRsingular}
(q_0-u_0)^2-(q_1-u_1)^2....-(q_{D-1}-u_{D-1})^2=\chi^2
\ee
This implies that the points of singularities lie on the hyperboloids. The
points $u_0,u_1,...u_{D-1},~\chi^2$ lie in the domain $\bf S$. The hyperboloids
where the parameters belong to $\bf S$ are called admissible. We have defined
a domain ${\bf{\bar R}}$, eq.(\ref{def-R}) where $F_C(q)$ is nonvanishing. Now
define a set $\bf R$ such that it is compliment to the real elements of
${\bf{\bar R}}$. Therefore, we conclude from the definition of ${\bf{\bar R}}$ that
$F_C(q)=0$ for every point which lies in $\bf R$ and is real. Moreover,
$F_R(q)=F_A(q)$ in this domain. This is the coincidence region. If we
examine the definition of domain ${\bf{\bar R}}$ (\ref{def-R}), it is bordered by
the upper branch of the hyperbola $(Q+q)^2={\cal M}_+^2$ and the other branch
is border of another hyperbola, $(Q-q)^2={\cal M}_-^2$. We arrive at the
conclusion that the region between these two hyperbolas can be identified as
the coincidence region. We recall that the set $\bf S$ is defined by the range of
values $u$ and $\chi^2$ take in the admissible parabolas. This set of values
is a subset of $(u,\chi^2)$ of all hyperbolas \cite{jl,sommer}.
We remind the reader that
the present discussion is for the case of equal masses. The more general
scenario follows from Dyson's analysis.\\
We are in a position to explore the analyticity properties of $F_R(q)$ since
we have defined various domain for our purpose. As stated earlier singularities
are in the admissible parabola defined by $(u-q)^2=\chi^2$.
It is better to, eventually, explore this feature in terms of invariants since scattering amplitudes
are expressed in terms of invariants. We focus on  the case of $Q\in V^+$.
Now we choose a frame where $Q=(Q_0,{\bf 0})$ where $\bf 0$ stands for
$(D-1)$-dimensional spatial components of the D-vector $Q$ in this frame. Next
we choose D-vector $q$  to find out the location of the singularities\footnote{
This treatment is analogous to that of \cite{jl,sommer} which is generalized to
D-dimensional spacetime}. In
order to simplify the calculations and bring out the essence of Jost-Lehmann
formalism, we make choice about $q$. We single out one spatial component of
$q$ and treat is as a variable to locate singularities and treat $q_0$ and
 and rest of the spatial components (now $D-2$ spatial coordinates are
fixed) as fixed parameters. The general case can
be treated more elegantly in Dyson's approach.
The above mentioned  choice would
lead us to find singularities in a simple way. To be specific let us choose
$q_1$ to be the variable of the $D-1$ spatial vectors; in other words
$q_0,q_2..q_{D-1}$ are treated as parameter and held fixed. If we examine
the Jost-Lehmann representation (\ref{jostlehFR}) then we note that we have
only Lorentz invariant objects appearing in the right hand side of the
equation. Therefore, with the present choice the of ${\bf q}^2$,
    the study of location of singularities is reduced to 
concentrating on $(q_1)^2$ as a variable. Thus
we are required to explore the location of the
singularities in the $q_1$-variable. These points are
\be
\label{q1-plane}
q_1=u_1\pm i
\sqrt{\chi_{min}^2(u)-(q_0-u_0)^2+(q_2-u_2)^2+...(q_{D-1}-u_{D-1})^2+\rho},~
\rho>0
\ee
Note that the set of points $\{u_0,u_1,..u_{D-1}; \chi_{min}^2=min~\chi^2 \}$
lie in the domain $\bf S$. We are able to identify the domain where the
singularities might reside with this choice for the variables $Q$ and $u$.
Another feature is  that the solution (\ref{q1-plane}) is
symmetric with respect to the real
axis. In general, the case is not so when masses are unequal;  the  original
derivation of Jost and Lehmann was applicable to equal mass case only. This
region contains all points in the region
${\rm Re}~q_1+i{\rm Im}~q_1,~{\rm Im}~q_1>0$ which satisfy
${\rm Re}~q_1+i{\rm Im}~q_1+\rho',~\rho'>0$.  In order to illustrate the result
in a very transparent manner, consider a very special case. Let us set
$q_0=q_2=...=q_{D-1}=0$, $u_1\ne 0$. Now the singular points are
\be
\label{q1-only}
 q_1=u_1\pm i\sqrt{\beta(u_1)^2+\rho,}~ \rho>0
\ee
and
\be
\beta^2(u_1)=Min~ \bigg[\chi_{min}^2-u_0^2+u_2^2+..+u_{D-1}^2 \bigg]~
 \ee
The minimization is taken over the variables $ \{u_0,u_2,...u_{D-1} \}$.
Now we  examine a still simpler case where the coincidence region is bounded by
two branches of hyperboloids (so that
${\cal M}_+^2={\cal M}_-^2={\cal M}^2$).
For this choice
\be
(Q+q)^2={\cal M}^2,~~~(Q-q)^2={\cal M}^2
\ee
and we have $\beta(u_1)={\cal{M}}-\sqrt{Q^2-u_1^2}$. \\
We briefly remark about Dyson's \cite{dyson}
 formulations of the problem to derive
representation for $F_C(q)$. This formalism is most suitable to obtain
representations for causal function and the retarded functions in any spacetime
dimensions. We present the details in the Appendix A.
 In order to obtain Jost-Lehmann representation,  in $D=4$, Dyson \cite{dyson} enlarged
the spacetime to six dimension with Minkowski signature such that there is
one time coordinate and five spatial coordinates. Therefore, there will be
six momentum variables as well. He constructed a six dimensional wave equation
in the momentum space and obtained solutions. Thus a causal function can be
defined in six dimensional spacetime and in terms of momentum space
variables. Then he chooses special set of coordinates and boundary conditions
to relate his solution to the function $F_C(q)$. Dyson made a very important
observation in relating  solution to his six dimensional wave equation with
the four dimensional causal function. He chose an arbitrary spacelike surface,
$\Sigma$, in the enlarged space. He used the fact that the solution can be
expressed in terms of its value and its normal derivative on an arbitrary
spacelike surface. Moreover, he demonstrated uniqueness of his solution.
We have generalized the technique of Dyson to arbitrary spacetime
dimension, D and derived the representation for the causal function, $F_C(q)$.
 A more general and
mathematically rigorous derivation of Dyson's theorem is presented in the
book of Bogolibov et. al. \cite{bogo} . Their starting point is to consider
 generalized functions in an n-manifold. Subsequently, they set up
the conditions for deriving the results of Dyson in a formal mathematical
frame work. 
It is possible that their formulation might be useful and more powerful 
in derivation of Dyson's theorem for   general D-dimensions.
The analyticity properties of the $2$-particle scattering amplitude and that
it is polynomially bounded have been rigorously derived by Hepp \cite{hepp}
in a more general setting without appealing to the LSZ formalism.
\bigskip

\noindent {\bf 3.3 D-dimensional Derivation of Lehmann's Ellipses}

\bigskip
\noindent
Our goal is to analyze the analyticity properties of the scattering amplitude,
$F(s,t)$ and write fixed-t dispersion relations. We have already argued in the
preceding section that for fixed negative $t$, as $s\rightarrow s_{thr}$,
$cos\theta$ exceeds its limit when $K^2\rightarrow 0$ in the above limit.
We had mentioned that Lehmann's investigation of analyticity of amplitude in
$t$ played an important role. Indeed, the Jost-Lehmann-Dyson representation
is very crucial for deriving Lehmann's result. We remind the reader the two
equations representing $F(s,t)$, (\ref{As}) and (\ref{Au}). Since we consider
scattering of equal mass identical particles, the kinematics is simplified.
We shall invoke the partial wave expansion for D-dimensional case in due
course (see Section 4).
 The following argument will convince that there is only one scattering
angle. Let us choose the c.m. frame system. Then D-momenta of particles, $a$
and $b$ are given by $p_a=(E_a, {\bf K})$ and $p_b=(E_b, -{\bf K})$ and
$E_a=\sqrt{m^2+K^2}=E_b$. Thus ${\bf K}$  defines a direction. For the out
going particles, c.m. momenta $|{\bf K}'|=|{\bf K}|$ 
in the equal mass scatterings. Thus
$p_c=(E_c, {\bf K}')$ and $p_d=(E_d,- {\bf K}')$, $E_c=E_d$. The two
 vectors $\bf K$ and ${\bf K}'$ (of equal magnitude) define a plane and
$cos\theta$ is the $cosine$ of angle between vectors $\bf K$ and ${\bf K}'$.
If we considered particles with spin, as it happens in the 4-dimensional case,
there will be more complications and the scattering amplitudes with
helicity states are to be defined. There will be analogous complications in
the case of D-dimensions when we consider scattering of particles with
spin. In fact we can choose a coordinate frame
to define the momenta of the four particles, now, 
\bea
\label{momenta}
&& p_a=({{\sqrt s}\over 2}, +K, 0, {\bf 0} ),~
p_b=({{\sqrt s}\over 2}, -K, 0, {\bf 0} )\nonumber\\&&
p_c=(-{{\sqrt s}\over 2},+Kcos\theta, +Ksin\theta, {\bf 0}),~
p_d=(-{{\sqrt s}\over 2}, -Kcos\theta, -Ksin\theta, {\bf 0})
\eea
We have adopted a coordinate system where $\bf K$ lies along spatial
direction '1' for $p_a$ and the same is true for $p_b$, except that sign is
opposite; all other components are zero. For the out going particles, the
c.m. momentum (magnitude $K$) lies along the $1-2$ plane. Thus $\bf 0$
represents a $D-3$ dimensional vector all whose components are zero. Moreover,
according to our conventions for scattering the vectors $p_c$ and $p_d$ have
appropriate signs. Furthermore,  $s=(p_a+p_b)^2=(-p_c-p_d)^2\ge 4m^2$,
$(p_a+p_b+p_c+p_d)=0$.  Now we identify
\be
\label{kinemat1}
q={1\over 2}(p_d-p_c)=(0, -Kcos\theta, -Ksin\theta, {\bf 0}),~~
P={1\over 2}(p_a+p_b)=({{\sqrt s}\over 2},0,0, {\bf 0})
\ee
The next step is to identify the coincidence region to proceed further. We
define (i) $ p_c^2=(P+q)^2< M_c^2$ and
(ii) $ p_d^2=(P-q)^2< M_d^2$. Where $ M_c$ is the lowest threshold
for two or more particle states which carry the same quantum number as $c$.
Similarly, $ M_d$ is defined for particle $d$. We keep these identities
for book keeping. However, for the present case
$ M_c^2= M_d^2 =4m^2$. We also need to use threshold constraints
in various channels. The relevant conditions are: (ia)
$ (p_a-p_c)^2 =(P-p_a+q)^2< {\cal M}_{ac}^2$ and (iib)
$(p_a+p_d)^2=(P-p_a-q)^2< {\cal M}_{ad}^2$. Notice that $ {\cal M}_{ac}$ and
${\cal M}_{ad}$ correspond to threshold for two or more particles carrying
quantum numbers of $a$-$c$ and $a$-$d$ channels respectively. Note, again that
${\cal M}_{ac}^2={\cal{M}}_{ad}^2=4m^2$ here as well. In general, the
problems arise when quantum number considerations forbid lowest two particle
state in a given channel. In the present case, the center of the hyperboloid,
$4m^2$ in the region $P+q\in V^+$ and same is true for the other one, i.e.
$P-q\in V^+$. We intend to find the location of the singularities. These are
in the region specified by the admissible hyperboloids i.e.
$(q-u)^2=\chi_{min}^2+\rho, ~\rho>0$ and $({(p_a+p_b)\over 2}\pm u)\in V^+$.
We determine $\chi_{min}^2$ from
\be
\label{chimin}
\chi_{min}^2=Max~\{0, {\cal M}-\sqrt{({(p_a+p_b)\over 2}+u)^2},
{\cal M}-\sqrt{({(p_a+p_b)\over 2}-u)^2} \}
\ee
We have an integral over $u$ in the Jost-Lehmann representation. We choose
\be
\label{u-choice}
u=(u_0, |{\bf u}|cos\alpha, |{\bf u}|sin\alpha, {\bf 0})
\ee
 and compute
$(q-u)^2$ with the constraint $(q-u)^2=\chi_{min}^2+\rho,~ \rho>0$.
This constraint leads to the equation            
\be
\label{u-eq}
u_0^2-{{\bf K}}^2-{\bf u}^2-2|{\bf K}||{\bf u}|cos(\theta-\alpha)=
\chi_{min}^2+\rho
,~\rho >0
\ee
Therefore, defining $cos(\theta-\alpha)= z(u)$,
\be
\label{z-eq}
z(u)={{1}\over{2|{\bf K}||{\bf u}|}}\bigg({\bf K}^2+{\bf u}^2+\chi_{min}^2+
\rho-u_0^2 \bigg)
\ee
Noting that $-1\le cos(\theta-\alpha)\le +1$. The singularity occurs for
$z_{min}=Min_{{\bf u},\chi^2}~z(u)$. It is found to be (for equal mass case)
\be
\label{z-min}
z_{min}(s)=\bigg(1+{{9m^4}\over{sK^2}} \bigg)^{1/2}
\ee
Note that the amplitude has singularities for $cos(\theta-\alpha)>z_{min}$.
Moreover, the amplitude is holomorphic in the interior of an ellipse in
the $cos\theta$ plane which has its focii located at $cos\theta=\pm1$,
with\footnote{see book of Itzykson and Zuber \cite{book3} for another method
to derive the Lehmann ellipse}
\be
\label{sle}
cos\theta=z_{min}cos\alpha+i\sqrt{z_{min}^2-1}~sin\alpha,~~~0\le\alpha\le2\pi
\ee
This is known as the small Lehmann ellipse (SLE) \cite{leh2}.
If we consider t-variable,
$t=2K^2cos\theta-2K^2$.  The domain of analyticity in the t-plane is
\be
-4K^2\le t\le 0
\ee
Thus it is quite satisfying  that the analyticity in the SLE is derived
for D-dimensional field theories from the axioms of LSZ.\\
We have remarked in the previous section that extension of the analyticity
domain to SLE is not quite adequate since as $K\rightarrow 0$,
$t\rightarrow \sim K$ and thus tends to zero softly \cite{book1}
compared to the earlier
case, before Lehmann proved existence of SLE. On that occasion, the amplitude
behave as $K^2$ near $s\rightarrow s_{thr}$ when it was assumed that
the amplitude is well defined in the region,
$-1\le cos\theta\le +1$.  However, with the existence of SLE, we are able
to go beyond.
 An important step was taken by Lehmann, in the $D=4$ theories,
 when he proved that
the absorptive part of the scattering amplitude is analytic inside a larger
ellipse - the large Lehmann ellipse (LLE). This is accomplished by reducing
the expression (\ref{As}) once more. In other words  two particles in state
$<-p_d~-p_c~out|$ were reduced and we were left with two particles in the state
$|p_a~p_b~in>$. If we reduce the latter
\bea
\label{4-reduced}
A_s(p_a...p_d)=&&2\pi\sum_n\int d^Dz'e^{-i{{(p_c-p_d)}\over 2}.z'}
<0|Rj_c({z'\over 2})
j_d(-{z'\over 2})|p_a+p_b,n>\times\nonumber\\&&
\int d^Dze^{-i{{(p_b-p_a)}\over 2}.z}<p_a+p_b,n|Rj_b^{\dagger}({z\over 2}))
j_a^{\dagger}(-{z\over 2})|0>
\eea
Notice that there is sum over physical 
intermediate states; however, the energy momentum
of the intermediate states is constrained to be $p_a+p_b$. We may apply J-L-D
construction to each of the matrix elements of the R-products. The
intermediate physical state is $|{\cal P}_n>=|p_a+p_b,n>$. The first matrix
element will be defined with a momentum and there will be an angle between
this vector and the momentum vector ${\bf{\cal P}}_n$. Similarly,
${\bf{\cal P}}_n$ will subtend an angle with the momentum vector associated
with second matrix element and we can choose this to be the c.m. frame
momentum vector ${\bf K}$ of $(p_a,p_b)$ system. We can implement J-L-D
construction to each of the two systems: (i) one system corresponds the initial
state $(p_a,p_b)$ and the other one is final state $(p_c,p_d) $ with c.m.
momentum ${\bf K}'$ (note $|{\bf K}|=|{\bf K}'|$). Moreover, the c.m. scattering
angle is inner product of the unit vectors along two directions;
 $cos\theta={\hat{\bf K}}.{\hat{\bf K}}'$.  Recall that in going through the
steps to derive SLE, we introduced an angle $\alpha$ in defining the
vector $\bf u$.
Now it requires two additional angular variables.
Moreover, if we denote the angle between $\bf K$ and ${\bf {\cal P}}_n$
as ${\cal{\phi}}_1$ then the angle between ${\bf{\cal P}}_n$ and ${\bf K}'$
is $\theta-{\cal \phi}_1$. It is easy to see that we need to introduce two
J-L-D functions, one 
for each matrix element: ${\bf \Phi}_1^n$ and ${\bf \Phi}_2^n$.
When we go through the algebraic steps, interestingly enough Lehmann's
method exactly goes through in this case also. The region of analyticity
of $A_s$ in $cos\theta$ plane is found to be an enlarged ellipse with focii
at $cos\theta=\pm 1$. The semimajor axis is $(2cos\theta_0^2-1)$ which is
$2z_{min}^2-1$ ; thus it is substantially larger compared to the semimajor axis
of the SLE.\\
Remarks: Now the amplitude is well defined for fixed t  as
$s\rightarrow s_{thr}$. As noted earlier, $A_s$ and $A_u$ are discontinuities
across the cuts $s>s_{thr}=4m^2$ and $u=4m^2-s-t>{\cal M}_{ac}^2=4m^2$.
Thus $A_s$ and $A_u$ are analytic in LLE. Therefore, we can write dispersion
relations in $s$ and $u$ variables for $A_s$ as well as for $A_u$ respectively.
Our goal in this investigation is not to prove dispersion relations; however,
we require knowledge of analyticity in $s$ and $t$ variables. We mention in
passing that all these results were obtained without utilizing the powers
of unitarity.\\
We proceed to investigate positivity conditions on the absorptive part $A_s$.
According to axiom (A1) the field operators act on a Hilbert space, $\cal H$,
to create states with positive norms. Let us envisage elastic scattering
and focus attention on $A_s$. We consider the scenario where the two particle
'out' state and the two particle 'in' have been reduced in the two step
process as we followed in deriving the LLE. We express (\ref{4-reduced}) in
a slightly different form
\bea
\label{positivity}
A_s(p_a,p_b,p_c,p_d)=&&
{\cal C}^2\int d^Dx_ad^Dx_bd^Dx_cd^Dx_de^{-ip_a.x_a-ip_b.x_b-ip_c.x_c-ip_d.x_d}
\nonumber\\&&
<0|R(j_c(x_c)j_d(x_d))R(j_b^{\dagger}(x_b)j_a^{\dagger}(x_a)|0>
\eea
It is understood that the total energy momentum conserving $\delta$-function
$\delta^D(p_a+p_b+p_c+p_d)$ multiplies  $A_s$ on the left hand side of the
 above equation and we have suppressed it. Here ${\cal C}$ is a real constant.
We are aware that the above matrix
element is a distribution. Therefore, it is necessary to introduce suitable
test functions in order that the right hand side is properly defined
\footnote{See discussion on this point by Martin in the ETH Lecture notes
\cite{book1}}. We
choose $C(x,y)$ to be such a test function which is rapidly decreasing and is
infinitely differentiable; the variables $x$ and $y$ belong to ${\bf R}^D$.
The Fourier transform of $C(x,y)$ is denoted as ${\tilde C}(p,q)$. Now
considered the smeared state
\be
\label{smeared}
\int d^Dx_ad^Dx_b C(x_a,x_b)R(j_b^{\dagger}(x_b)j_a^{\dagger}(x_a)|0>
\ee
This state has a positive norm and therefore,
\bea
\label{norm}
&& 0<||\int d^Dx_ad^Dx_bC(x_a,x_b)R(j_b^{\dagger}(x_b)j_a^{\dagger}(x_a)|0>||^2
\nonumber\\&&
=<0|\int d^Dx_cd^Dx_dC^{*}(x_c,x_d)R((j_c(x_c)j_d(x_d)) \times\nonumber\\&&
\int  d^Dx_ad^Dx_b C(x_a,x_b)R(j_b^{\dagger}(x_b)j_a^{\dagger}(x_a)|0>
\eea
Let us take the Fourier transform of (\ref{smeared})
\bea
\label{FT-smeared}
0<&& {\cal C}'^2\int d^Dp_ad^Dp_bd^Dp_cd^Dp_d{\tilde C}(p_a,p_b){
\tilde C}^{*}(p_c,p_d)\nonumber\\&&
\delta^D(p_a+p_b+p_c+p_d)A_s(p_a,p_b,p_c,p_d)
\eea
This integral is positive as defined and ${\cal C}'$ is a real constant.
Now define
\bea
\label{kin1}
P={({p_a+p_b)}\over 2},~~~q=-{{(p_a-p_b)}\over 2},~~~q'={{(p_c-p_d)}\over 2}
\eea
Thus we get the relations (i) $p_a=P+q$, $p_b=P-q$ and (ii)
$p_c=-P+q', p_d=-P-q'$; using $\sum p_l=0, l=a,b,c,d$. Moreover, Mandelstam
variables $s$ and $t$ can be expressed in terms of $P,q,q'$ also. We choose
${\tilde C}(p_a,p_b)={\tilde C} (P+q, P-q)={\tilde f}(P){\tilde g}(q)$
The functions, $\tilde f$ and $\tilde g$, defined in the momentum space are in 
$R^D$. The positivity condition (\ref{FT-smeared}) is expressed as    
\bea
\label{new-pos}
\int d^DP|{\tilde f}(P)|^2\int d^Dqd^Dq'A_s(-P-q,-P+q,-P-q',-P+q'){\tilde g}(q)
{\tilde g}^{*}(q') >0
\eea
We note that $A_s$ is a positive measure in $P$. The expression for $A_s$ in
(\ref{positivity}) which is defined in the coordinate space is product of
two retarded functions. One is $Rj_c(x_c)j_d(x_d)$ and other is
$Rj^{\dagger}_b(x_b)j^{\dagger}_a(x_a)$ Thus each of the R-product satisfies
retardedness properties in pair of coordinates: the former in $(x_c,x_d)$ and
the latter in $(x_a,x_b)$. We argue that if $P$ is held fixed, then $A_s$
is a function of $q$ and $q'$ and it possesses analytic properties in the two
variables. We arrive at the conclusion that the domain of holomorphy (for $q$)
lies in a domain, $\Sigma \in C^D$ (the space of complex coordinates) and the
other variable ($q'$) is also in the same domain. We remind the reader the
steps we followed in the context of deriving J-L-D representation for
the scattering amplitude. First we obtained the domain of analyticity for
$F(s,t)$ itself and, in the next step, we derived the larger analyticity
domain for the absorptive part of the amplitude, $A_s$. This was the route
taken   to arrive at LLE. In the present context,let us take $P$ to be real
and hold it fixed. Consequently, the positivity conditions
\bea
\label{measures}
\int d\mu(q)d\mu(q')A_s(-P-q,-P+q,-P-q', -P+q') \sigma(q)\sigma(q')\ge 0
\eea
The measures are: $d\mu(q)=dq\wedge dq^{*}$ and  $d\mu(q')=dq'\wedge dq'^{*}$.
Moreover, $q,q'$ are in $C^D$ and $\sigma(q)$ is a suitably defined function
in $L_2$. Now on we suppress the presence of test functions since we know
that the distributions are defined with them; we might explicitly invoke
their presence if necessary.  We choose the following assignments
for various D-vectors which are useful for our kinamatical analysis and
to study consequences of positivity.
\bea
\label{kin2}
&&P=({\sqrt s},0,0, {\bf 0}),~q=({\sqrt{K^2+m^2}}-\sqrt{s},Kcos\phi,Ksin\phi,
{\bf 0})\nonumber\\&&~
q'=({\sqrt{K^2+m^2}}-\sqrt{s},Kcos\phi',Ksin\phi',
{\bf 0}),
\eea
Note that as before, $\bf 0$ is the $D-3$ dimensional vector of the
spatial vector with $D-1$ components. We can get expressions for $s$ and
$t$ with the above assignments for $P, q,q'$. Of special interests to us is
the vector $(q-q')$ since $t=(q-q')^2$. Note that
\be
(q-q')=(0,K(cos\phi-cos\phi'), K(sin\phi-sin\phi'), {\bf 0})
\ee
Thus $t=-2K^2[1-cos(\phi-\phi')]$. Therefore, we may identify $\phi-\phi'=\theta$ and
$cos(\phi-\phi') = cos\theta$. Now the scattering amplitude is a function of
$s$ and $cos(\phi-\phi')=cos\theta$. In the case of $D=4$ theories it was
convenient to choose  ${\tilde g}(q)=e^{im\phi}$ and
${\tilde g}^{*}(q')= e^{-im\phi'}$ then the positivity property of
$A_s^{D=4}(s,t)$ could be proved since the amplitude is expanded in the
Legendre polynomial basis \cite{book1,book2}.\\
In the case of scattering in $D$-dimensions, the basis functions are the 
Gegenbauer polynomials as has been remarked earlier. In this case the
positivity property holds also. We shall show positivity of $A_s$ when
we study the partial wave expansion and the problem of enlarging the  domain
of analyticity. The reader may consult the Appendix C  where we have
collected some useful formulas relevant in our work where the Gegenbauer
 polynomials appear.

\bigskip

\noindent{ \bf 3.4 Fixed t Dispersion Relations}

\bigskip
\noindent Let us discuss the fixed-t dispersion relations in $s$ for the
scattering amplitudes. We postulated that there are no bound states and
therefore, $s_{thr}=s_{phys}=4m^2$. The elastic scattering amplitude,
for fixed $t$ in the region, $-T\let\le 0, T>0$, admits
a dispersion relation in $s$. If the integrand in the dispersion relation does
not have a good convergence property then an unsubtracted dispersion relation
may  be substituted  by a subtracted dispersion relation with $N$ subtractions.
We know from our earlier discussions that in the LSZ formulation $N$ is finite.
 Moreover, $F(s,t)$ is also an analytic function in both $s$
and $t$ in some neighborhood of any ${\tilde s}$ where ${\tilde s}$ lies
in some interval below $s_{phys}$ such that $4m^2-\delta<{\tilde s},4m^2$ and
it also lies in some neighborhood $|t|<{{\tilde R}}({\tilde s})$ of $t=0$.
This result has been proved for the case of $D=4$ in \cite{beg1}.
Lehmann \cite{leh3}  (in $D=4$) approached the problem of
writing fixed-t dispersion relation from a different view point. He considered
scattering amplitude for fixed $s>s_{thr}$, for values of $t$ that lie within
SLE and thus $t$-analyticity is valid. This is to be contrasted with results of
\cite{beg1}. As we shall see later, Martin exploited the analyticity domain
 ordained in \cite{beg1} to prove his theorem.
 We have not rigorously proved existence of such a domain of analyticity
for D-dimensional theories. This analyticity property was derived  by
\cite{beg1}  in the LSZ formulation,
with micro causality. 
 We have shown that there is analyticity in $t$ in the Lehmann ellipse
for D-dimensional theories and polynomial bounded (in $s$)
in such higher dimensional theories. Therefore, we strongly believe that BEG
result is also valid for D-dimensional theories (see discussion (iv) after
(\ref{edge})). The absorptive parts
$A_s$ and $A_u$ defined on the right hand and left hand cuts respectively, for
$s'>s_{thr}$ and $u'>u_{thr}$ are holomorphic in the LLE. Thus, assuming no
subtractions necessary
\be
\label{disperse}
F(s,t)={1\over \pi}\int _{4m^2}^{\infty}{{ds'~A_s(s',t)}\over{s'-s}}
+{1\over \pi}\int _{4m^2}^{\infty}{{du'~A_u(u',t)}\over{u'+s}-4m^2+t}
\ee
We have argued that the integrands satisfy polynomial boundedness properties
and therefor, if required, we might need subtractions.

\newpage

\noindent{\bf 4. Analyticity in $s$ and $t$ and the Asymptotic Behavior of
Scattering Amplitude in D-dimensions}

\bigskip
\noindent
We shall further study the analyticity properties of the
scattering amplitude in this section. Let us recall the main results of
the previous sections. We have derived the expressions for the absorptive
parts of the scattering amplitude $A_s$ and $A_u$ from LSZ formulation and we
have argued that the amplitude is polynomially bounded in $s$ when we write
a dispersion relation. In Section 3, we devoted our investigations to
the analyticity properties in that the representations for $F_C(q)$ and
$F_A(q)$ were obtained from the Jost-Lehmann-Dyson theorem. We also showed
that in D-dimensions, the analog of Lehmann ellipses exist. However, it was
mentioned that the results of generalized Lehmann ellipses were not adequate
to derive the Froissart bound. \\
In order to derive the higher dimensional Froissart-Martin bound we have
to go through a few more important steps. This will be focus of this section.
A crucial result in the  derivation of the Froissart-Martin bound,
as is known in the present form, relies on a theorem due to Martin \cite{am}. We shall
present generalized Martin's theorem. However, we shall summarize below
the essence of the theorem, as was derived in the four dimensional case
and provide remarks on our way to generalize the theorem.\\
 {\it Statement of Martin's Theorem for $D=4$}: If following
requirements are satisfied\\
I. $F(s,t)$ satisfies fixed-t dispersion relation in s with finite number of
subtractions ($-T\le t\le 0$).\\
{\it Remark: This property is true for the case of D-dimensional theory in
LSZ formulation as has been argued by us.}\\
II. $ F(s,t)$ is an analytic function of the two Mandelstam variables, $s$ and
$t$, in a neighborhood of $\bar s$ in an interval below the threshold,
$4m^2-\rho<{\bar s}<4m^2$ and also in some neighborhood of $t=0$,
$|t|<R({\bar s})$. This statement hold due to the work of
Bros, Epstein and Glaser \cite{beg1,beg2}.\\
{\it Remark: We have not proved the BEG theorem for the four point
amplitude in the D-dimensional case.
However, there seems to be no serious obstacles to generalize it to D-dimension in 
the LSZ formalism. Thus if we follow the arguments presented following equation 
(\ref{edge}) (see the remark (iv)) the BEG results
 hold for the four point amplitude.
}\\
III. Holomorphicity of $A_s(s',t)$ and $A_u(u',t)$: The absorptive parts of
$F(s,t)$ on the right hand and left hand cuts with $s'>4m^2$ and $u'>4m^2$
are holomorthic in the LLE. \\
{\it Remark: We have shown that  for D-dimensional field theories, there
exist SLE and LLE (see Sec 3.3).
 This result was derived on the basis of the generalized
J-L-D construction in the D-dimensional case.}\\
IV. The absorptive parts $A_s(s',t)$ and $A_u(u',t)$, for $s'>4m^2$ and
 $u'>4m^2$ satisfy the following positivity properties
\bea
\label{positivityI}
\bigg|{\bigg({\partial\over{\partial t}}}A_s(s',t)\bigg)^n\bigg|
\le {\bigg({\partial\over{\partial t}}\bigg)^n} A_s(s',t)\bigg|_{t=0},~~
-4K^2\le t\le 0
\eea
and
\bea
\label{positivityII}
\bigg|{\bigg({\partial\over{\partial t}}}A_u(u',t)\bigg)^n\bigg|
\le {\bigg({\partial\over{\partial t}}\bigg)^n} A_u(u',t)\bigg|_{t=0},~~
-4K^2\le t\le 0
\eea
{\it Remarks: The above positivity properties were proved, in $D=4$ case,
using properties of the Legendre polynomials in the partial wave
expansion for the scattering amplitude with unitarity constraints
on the partial wave amplitudes. For the D-dimensional case, the basis
functions are the Gegenbauer polynomials. We shall prove similar inequalities
for absorptive parts in the case of D-dimensions in the latter part of
this section. The inequalities (\ref{positivityI}) and (\ref{positivityII})
indeed hold.}\\
The Martin's theorem aims at deriving the domain of analyticity of
the scattering amplitude
in the complex planes of $s$ and $t$ variables. This can be proved if the
scattering amplitude can be expanded in a power series
\be
\label{martin1}
F(s,t)=\sum_{n=0}^{\infty} {{t^n}\over{n!}}{({d\over{dt}})^n}F(s,0),
\ee
Then appealing to the Hartog's theorem\footnote{Hartog's theorem for functions
of several complex variables $ f(z_1,z_2,..z_n)$ may stated as follows. Let
$ f(z_1,z_2,..z_n)$ be defined in an n-complex domain $D$ and let
$ f(z'_1,z'_2,z_k,z'_{k+1}..z'_n)$ for all $k=1,2,...n$ be holomorphic in
$|z_k-z'_k| <\epsilon$ as a function of $z_k$ for all
$ f(z'_1,z'_2,z'_k,z'_{k+1}..z'_n)\in D$. Then $ f(z_1,z_2,..z_n)$ is
holomorphic in $D$ simultaneously. See \cite{book1} for examples to illustrate
the technique of analytic continuations and how the domain of holomorphicity
is enlarged.}
, $F(s,t)$ will be analytic in the
quasi-product of topological domains
 $D_s\otimes D_t$ if for every $s\in D_s$ the
series is uniformly convergent for $t\in D_t$.
Therefore, the first step is to define and derive a bound on
${({d\over{dt}})^nF(s,t)}|_{t=0}$ and then analyze the convergence properties
of (\ref{martin1}). Let us first consider the dispersion relation for $F(s,t)$
only in the presence of 
the right hand cut and momentarily ignore the presence of
the left hand cut. We shall account for the presence of the left hand cut later.
Moreover, we also assume that there is no subtraction.
\be
\label{martin2}
F(s,t)={1\over \pi}\int_{s_{thr}}^{\infty}ds'{{A_s(s',t)}\over {s'-s}}
\ee
Next, we use the BEG \cite{beg1,beg2}
result that for each point of $s$ in the cut plane and
fixed $t$, it is possible to write a dispersion relation. More specifically,
if one chooses $s=4m^2$, there is an analyticity neighborhood in
$s$ and in $t$ such that ${\bar s}=4m^2-\epsilon$ and $t$ in the
analyticity neighborhood. One can choose an $\epsilon$ which is small enough
to fulfill these requirements. Thus when we have chosen an $\bar s$ which
is sufficiently close to $s_{thr}$, we can find an $R$ such that
$F({\bar s},t)$ is analytic in $t$ in the region $|t|<R$. We need to choose
$\bar s$ below the threshold to avoid the divergence difficulties of the
integral above (\ref{martin2}). Therefore, we rewrite the above relation as
\be
\label{martin3}
F({\bar s},t)={1\over \pi}\int_{s_{thr}}^{\infty}ds'{{A_s(s',t)}\over {s'-
{\bar s}}}
\ee
Notice that $F({\bar s}, t)$ is analytic in the domain $|t|<R$. Consequently,
all derivatives of the function exists at $t=0$. Now if we apply Cauchy's
inequality \cite{tit} for an analytic function inside a domain
$|t|<R-\epsilon$ then its derivatives at $t=0$ are bounded  i.e.
\be
\label{martin4}
|F({\bar s},t)|\le M,~~~{\rm and},~~~
|({d\over{dt}})^nF({\bar s},t)|<{{n!M}\over{(R-\epsilon)^n}}
\ee
We follow Martin's argument and endeavor to show that, if we adopt his
procedures, there is a method to analytically continue the inequalities
(\ref{martin4}) to the complex cut $s$-plane. This is accomplished by utilizing
the powerful result of fixed-t dispersion relations. The above result can be
proved if one is permitted to interchange differentiation with respect to $t$
with the integration over $s$ in the dispersion relation. The Lemma 1, given
in the Appendix B can be readily utilized for this purpose. Thus
\bea
\label{martin5}
{d\over{dt}}F({\bar s},0)=&&{\rm lim}_{\tau\rightarrow 0,\tau>0}~
{{F({\bar s},0)-F({\bar s},-\tau)}\over\tau}\nonumber\\&&
={1\over\pi}{\rm lim}_{\tau\rightarrow 0,\tau>0}\int_{s_{thr}}^{\infty}
{{A_s({\bar s},0)-A_s({\bar s}-\tau)}\over{\tau(s'-{\bar s})}}ds'
\eea
The absorptive part $A_s({\bar s}, t)$ and its t-derivatives satisfy
 the inequalities
\bea
\label{martin6}
&& A_s(s, t=0)\ge\bigg|A_s(s, t) \bigg|_{-4k^2\le t\le 0} \nonumber\\&&
{\bigg({d\over{dt}}\bigg)^n}A_s(s,t)\bigg|_{t=0}\ge\bigg|
\bigg({d\over{dt}}\bigg)^n A_s(s,t)\bigg|_{-4k^2\le t\le 0}
\eea
The first inequality follows from optical theorem applied to elastic amplitude
using positivity property of the imaginary part of the  
partial wave amplitudes. As noted, for D-dimensional case, the basis
functions are the Gegenbauer polynomials and the first inequality is easily
derived. The other inequalities need some careful analysis to use relations
among the Gegenbauer polynomials. At this moment, we would like the reader to
accept (\ref{martin6}) and see the subsequent results in this section. We
use the arguments of Lemma 1 and decompose the right hand side of the
integral eq. (\ref{martin5}) as $\int_{s_{thr}}^X+\int_X^{\infty}$ then we
argue as in Lemma1 that
\be
\label{martin7}
{\rm lim}_{\tau\rightarrow 0}~{{F({\bar s},0)-F({\bar s}-\tau)}\over{\tau}}\ge
{\rm lim}_{\tau\rightarrow 0}~\int_{s_{thr}}^X{{A_s({\bar s},0)-
A_s({\bar s},-\tau)}\over{ \tau(s'-{\bar s})}}
\ee
as is obvious. We are free to choose $\tau$ to be small enough so that for all
$4m^2\le s'\le X$ the interval $-\tau\le t\le 0$ inside all the Lehmann
ellipses for a given $X$. We remind the reader that the size of the Lehmann
ellipses depend on the value of $s'$ we choose since the semimajor axis
depends on $s'$.  Thus
\be
\label{martin8}
{\rm lim}_{\tau\rightarrow 0}~{{A_s(s',0)-A_s(s',-\tau)}\over\tau}=
{{dA_s(s',0)}\over{dt}}
\ee
and using Lemma 1
\be
\label{martin9}
{{dF(\bar{s},0)}\over{dt}}\ge{1\over\pi}\int_{s_{thr}}^{\infty}
{{dA_s(s',0}\over{dt}}{{1}\over{s'-{\bar s}}}ds'
\ee
Our aim is to show that $F({\bar s}, t)$ can be expanded in a power series.
Thus we should prove the inequality other way around for $t$-derivative
of the amplitude. To this end, let us begin with    
\be
\label{martin10}
{{dF({\bar s}, 0)}\over{dt}}={1\over\pi}{\rm lim}_{\tau\rightarrow 0}~
\int_{s_{thr}}^{\infty}ds'{{dA_s(s',-\tau'(s'))\over{dt}}}{1\over{(s'-{\bar s})}}
\ee
Now choose $-\tau<-\tau'(s')<0$ (it corresponds to a physical region).
Moreover, if we have $s'>s_{thr}+\rho, \rho>0$, $ t=-\tau'(s')$ is in physical
region. We can utilize the positivity property and the inequality
associated with the absorptive part
\be
\label{martin11}
{{dF({\bar s},0)}\over{dt}}\le{\rm lim}_{\tau\rightarrow 0}~{1\over\pi}
\int_{s_{thr}}^{{s_{thr}}+\rho}ds'{{dA_s(s',-\tau'(s'))}\over{dt}}
{1\over{s'-{\bar s}}}
+\int_{{s_{thr}}+\rho}^{\infty}ds'{{dA_s(s',0)}\over{dt}}{1\over{s'-{\bar s}}}
\ee
The point to notice that in the above equation the first integral is taken over
a compact interval. Moreover, the integrand is positive and regular. Therefore,
we are permitted to choose $\rho$ as small as we desire. As a consequence,
\be
\label{martin12}
{{dF({\bar s},0)}\over{dt}}\le {1\over\pi}\int_{s_{thr}}^{\infty}
{{ds'}\over{s-s'}}({{dA_s(s',0)}\over{dt}})
\ee
Thus combining (\ref{martin9}) and (\ref{martin12}) we come to conclusion that
that
\be
\label{martin13}
{{dF({\bar s},t)}\over{dt}}\bigg|_{t=0} ={1\over\pi}\int_{s_{thr}}^{\infty}
\bigg({{A_s(s',0)}\over{dt}}\bigg){{ds'}\over{(s'-{\bar s})}}
\ee
Therefore, using the above result for
the t-derivative of $F(s,0)$ we can derive the relation
\be
\label{martin14}
{\bigg({d\over{dt}}\bigg)^n}F({\bar s},t)={1\over\pi}\int_{s_{thr}}^{\infty}
{{{({{d}\over{dt}})^n}A_s(s',t)}\over{s'-{\bar s}}}
\le{{Mn!}\over{(R-\epsilon)^n}}
\ee
 Note that
the positivity property   i.e.
${({d\over{dt}})^n}A_s(s',0)\ge 0$ has been utilized.  
We shall demonstrate it  later in this
section. Thus
 the absorptive part is also bounded.
The last inequality is a consequence of the Cauchy's inequality. We can also
define
\be
\label{martin15}
{\bigg({d\over{dt}}\bigg)^n}F(s,0)={1\over\pi}\int_{s_{thr}}^{\infty}
{{({d\over{dt}})^nA_s(s',0)}\over{s'-s}}ds'
\ee
The function
as defined in (\ref{martin14}) is defined in a finite segment of the real
s-axis. We are allowed to move $\bar s$ and the argument does not change.
If the integral has uniform convergence, we can continue to it arbitrary complex
plane. Thus  ${({d\over{dt}})^n}F(s,0)$ can be continued to arbitrary complex
$s$-plane since there exist an analyticity neighborhoods. Since
${({d\over{dt}})^n}A_s(s,t)|_{t=0}\ge{({d\over{dt}})^n}
A_s(s,t)|_{-4K^2\le t\le 0}$, we use
\be
\label{martin16}
\bigg|{{{({d\over{dt}})^n}A_s(s',0)}\over{s'-s}}\bigg|<
{{{({d\over{dt}})^n}A_s(s',0)}\over{s'-s}}\mu(s,{\bar s})
\ee
and
\be
\label{martin17}
\mu(s,{\bar s})=~\rm {sup}_{4m^2<s'<\infty}~\bigg|{{s'-{\bar s}}\over{s'-s}}
\bigg|
\ee
note that this remains finite as long we are outside the cut. Moreover,
for ${\rm Re~s}>s_{thr}$ as ${\rm Im s}\rightarrow 0$, $\mu(s,{\bar s})$
diverges. We utilize the
earlier inequality to argue that
\be
\label{martin18}
\bigg|{({d\over{dt}})^n}F(s,0)\bigg|<\mu(s,{\bar s}){{Mn!}\over{(R-\epsilon)^n}}
\ee
Thus the expansion of $F(s,t)$ in a power series in $t$, (\ref{martin1}),
 converges for $|t|<R$. As a consequence we conclude that, for fixed $s$,
 $F(s,t)$ is an analytic function of $t$. Moreover, for any fixed,
$t$, $|t|<R$, if we remain in a compact region of complex $s$-plane where
$\mu(s,{\bar s})$ is bounded; it converges. Note that each term in the power
series expansion of $F(s,t)$ is analytic in $s$. It follows from Hartog's
\cite{hert,book1}
theorem that the amplitude is analytic in topological product of the domains
$D_s\otimes D_t$. This is defined by: $|t|<R$ and s outside the cut
$s_{thr}+\lambda=4m^2+\lambda, \lambda>0$.\\
The importance of this result is recognized if we recall 
BEG theorem \cite{beg1}.
It was shown that in neighborhood of any point $s_0,t_0$, $-T<t_0\le 0$,
$s_0$ outside the cuts, there is analyticity in $s$ and $t$ in a region
\be
\label{martin18a}
|s-s_0|<\eta(s_0,t_0),~~~~~|t-t_0|<\eta(s_0,t_0)
\ee
The size of this analyticity neighborhood can vary as we vary $s_0$ and
$t_0$.
 Moreover,
as $s\rightarrow 0$, $\eta(s)$ can shrink to zero. Martin's theorem
proves that there is a lower bound to $\eta(s)$ such that $\eta(s)\ge R$ and
$R$ is independent of $s$. Consequently, we can argue that in the region
$|t|<R$, $F(s,t)$ satisfies unsubracted dispersion relation; that is where
we started with.  Notice that if the amplitude satisfies polynomial
boundedness, then with subtractions, the analyticity properties are not
affected.\\
We remind the reader that the above result was derived in the
presence of only the  right hand
cut. There is a problem, when we include the left hand cut part,
$A_u(u',t)$. Let us recall
\be
\label{martin19}
F({\bar s},t)={1\over\pi}\int_{s_{thr}}^{\infty}ds'{{A_s(s',t)}
\over{(s'-{\bar s})}}+{1\over\pi}\int_{u_{thr}}^{\infty}
{{A_u(u',t)}\over{(u'-{\bar s})}}
\ee
and ${\bar u}=4m^2-{\bar s}-t$. The following problem crops up when we take
into account the presence of the left hand cut in the dispersion relation.
If we take derivative of $F({\bar s},t)$ at $t=0$, then right hand side of
(\ref{martin19}) is not a sum of two positive terms (and positivity played
a key role in all preceding arguments in this context). The positivity
property is spoiled due to the presence of ${\bar u}=4m^2-{\bar s}-t$ in
the denominator of the second integral and its presence makes the previous
procedure inadequate. The Lemma 2 is very useful to resolve this issue (see
Martin's lecture notes for details)\footnote{See the works of Martin
\cite{book1} and also see \cite{sommer,sommer1}}. 
The analyticity properties of
$F({\bar s},t)$, in both $s$ and $t$ was proved in the presence of only the
right hand cut.  If we consider a function
\be
\label{martin20}
{{F({\bar s},t)}\over{{\bar{s}-R-t}}}
\ee
it is also analytic in $|t|<R$. The analyticity property of newly defined
function (\ref{martin20}) is unaffected 
as long as $R<{\bar s}<s_{thr}$. In order to
facilitate application of Lemma 2, we identify
\be
\label{martin21}
G_1(s',t)=A_s(s',t),~~~G_2(s',t)={{1\over{(s'-{\bar s})}}}
{{1\over{(s'-R-t)}}},~~\beta=s_{thr}
\ee
for the right hand cut contribution to dispersion integral. For the left hand
cut now our identifications are
\bea
\label{martin22}
&&G_1(u',t)=A_u(u',t)\nonumber\\&&
G_2(u',t)={{1\over{(u'-{\bar s}-4m^2+t)}}}
{{1\over{(s'-R-t)}}},~z=-s'+4m^2-t,~\beta=u_{thr}
\eea
The functions identified with $G_1$ and $G_2$ defined in Lemma 2 satisfying the
requirements (b) and (c) laid down in proving Lemma 1 and Lemma 2. Therefore,
     in going through Martin's arguments for the function in complex
$s$-plane.  The new bound is
\bea
\label{martin23}
\bigg|F(s,t)\bigg|<&&\bigg[{\rm sup}_{s_{thr}\le s'\le\infty}
\bigg|{{s'-{\bar s}}
\over{{s'-s}}}\bigg|+{\rm sup}_{u_{thr}\le u'\le\infty}
\bigg|{{u'-{\bar s}-4m^2}\over{{u'+s-4m^2+t}}}\bigg|\bigg]\times\nonumber\\&&
({\bar s}-R)\sum_{n=0}^{\infty}{{|t|^n}\over{{R^n}}}{\rm Max}_{|t|<R}
\bigg({{F({\bar s},t)}\over{{{\bar s}-R-t}}}\bigg)
\eea

\bigskip
\noindent{\it Determination of R:} We remark that in derivation
of the generalized 
Martin's theorem, it was crucial to to use the fact that the power series
expansion of the scattering amplitude, $F(s,t)$ converges in the domain
$|t|<R$. The second important point to note is that the BEG function,
$\eta(s)>R$, not only it is bounded from below but $R$ is independent of s.
Therefore, it is essential to determine R. We also recall that in the
preceding discussion, we have used the fact that $F(s,t)$ in analytic
in the SLE whose existence was derived in Section 3. Moreover, the analyticity
of $A_s$ and $A_u$ inside the LLE is also a very important ingredient in
writing the dispersion relation. Note that the right hand extremity of
SLE is $r_{sLE}=2K^2(z_{min}-1)$ whereas that of the LLE is
$r_{lLE}=4K^2(z_{min}^2-1)$. \\
We recall that the generalized Martin's theorem was derived for
${\bar s}<s_{thr}$, it lies just below the threshold although it lies in
the analyticity region as per of BEG results.
As has been pointed out earlier, BEG formalism
is not adequate to determine $R$. Therefore, it is important to determine
$R$ and to employ the analyticity in $t$ to arbitrary $\bar s$, not just
below the threshold. Thus, it is worth while to utilize Lehmann's analysis
which implies that if $s_1>s_{thr}$, then $F(s_1,t)$ is analytic within the
domain
\be
\label{martin24}
|t|<2K^2(s_1)\bigg[\bigg(1+{{({\cal M}_a^2-m_a^2)({\cal M}_b^2-m_b^2)}
\over{{K^2(s_1)[s_1-({\cal M}_a-{\cal M}_b)^2]}}}\bigg)^{1/2}-1\bigg]
\ee
A few comments are necessary: (i) Note that c.m. momentum $K=K(s_1)$ and
$s_1$ is above the physical threshold. (ii) We
have deliberately retained ${ M}_a, { M}_b, m_a, m_b$ and remind that
$({M}_a, { M}_b)$ correspond to two or more particle states carrying
the quantum number of 'a' and 'b'. In our case these are equal and
$({ M}_a= { M}_b=2m)$ since $m_a=m_b=m$. In case of $\pi$-$\pi$
scattering it starts with $3\pi$ state due to $G$-parity considerations. (iii)
$s_1$ is on the physical cut. The analytic continuation mentioned earlier
is applicable in the range $0<s={\bar s}<4m^2$. Sommer \cite{sommer1}
provided a general method to remove the cut $4m^2<s<s_1$
which permits to obtain
Martin's result beyond ${\bar s}$. This procedure led to derivation of
 the value of
$R$.\footnote{ Martin \cite{am} had derived $R=4m_{\pi}^2$
for $\pi\pi$ scattering.} Martin first applied the technique, described below,
  to obtain value of $R$ 
for $\pi^0\pi^0$ scattering.  Martin's
original arguments for $\pi^0\pi^0$ scattering is applicable for the 
case at hand since
the essential logical sequences are the same. 
We consider scattering of equal mass 
neutral spinless particles. Therefore, all the three channels,
$s$, $t$ and $u$,  are identical and
consequently, Martin's procedure goes through. 
 Then following the BEG \cite{beg2} argument
the dispersion relation for fixed t ($-4K^2<t\le 0$) can be written down. 
Consequently,  all the points  inside
 the region (a triangle): (a) $-4m^2<t \le 0$, $s< 4m^2$, $u<4m^2$ lie in the
domain of analyticity for the scattering amplitude. Note that $t$ can take
a much lower value; however, we have chosen a value which is required
for what follows in sequel.  
We can also consider the
other two channels:  the one choice of kinematical region is (b)
$-4m^2<s\le 0$, $t<4m^2$, $u<4m^2$ and the other is (c)
$-4m^2<u\le 0$, $t<4m^2$, $s<4m^2$.\\  
As in the case of region (a), for the two other  regions,
 (b) and (c), to each of
the points in (b) as well as in (c), 
we can attach a neighborhood in $s$ and $t$ 
where the scattering
amplitude is analytic. Note that if we choose $s=s_1$, there is an analyticity
domain $|t|<R$. Moreover, the same analyticity argument goes through for
$s_1<s<4m^2$. Now fix $s$, $4m^2-R+\epsilon <s<4m^2$; 
following \cite{beg2}, for each $-s<t_0<4m^2$ there is a neighborhood
$|t-t_0|<\eta(s,t_0)$ of analyticity in $t$. Note that we have analyticity in
this compact region. Then invoke the Heine-Borel-Lebesgue theorem to argue
that the interval $-s+\epsilon\le t\le 4m^2-\epsilon$ can be covered
by finite number of (such) compact intervals. Moreover, the fixed-$s$ amplitude
is analytic in t in a region which also contains the real domain
$-s+\epsilon\le t<4m^2-\epsilon$. First consider the case when an 
unsubtracted dispersion relation can be written for $F(s,t)$ for 
$0<s<4m^2$. If we consider a function
\be
\label{triangle}
{{F(s,t)}\over{s-t}}
\ee
in the interval $0<s<4m^2$, then the function has all its $t$-derivatives
positive at $t=0$. The power series expansion of this function acquires a
singularity at $|t|=R$. Thus, the strip $-s+\epsilon<t<s+\epsilon$
is singularity free. Therefore, one  concludes $R=s$. Now $s$ can be taken
as much closer to $4m^2$ as we desire. Therefore, the analytic 
domain is $|t|<4m^2$. \\
Remarks: (i) If the amplitude needs subtractions to write dispersion relation
then it allows Martin's arguments to go through.  We can construct suitable
amplitude in this case and implement Martin's prescription to determine $R$. 
Note that the the dispersion relation still holds for the newly defined
amplitude \cite{am}.
(ii) The above arguments of Martin goes through for the elastic
amplitude for scattering of massive neutral scalar particles in the
D-dimensional theories as has been argued above. The crucial point
to underline here is that all three channels are identical. Therefore, 
the amplitude $F(s,t)$ satisfies all the criterion we require to determine
$R$.\\
(iii) The contrasting point is that in 
the case of hadronic scattering (i.e. $\pi\pi$
scattering)
the  of Martin's triangle is determined through the 
introduction of $m_{\pi}$ which is experimentally determined to have a
numerical value i.e. $m_{\pi}=140$ MeV. This value is not determined from
the theory but it is experimentally measurable as a number. 
However, what will be $R$ (that it is
$4m_{\pi}^2 -\epsilon$) is derived from the theory.
 What is important is that value of $R$ is expressed
in terms of the mass parameter in the theory. In the case of scattering
in D-dimensional theory, $R$ is also determined in terms of the mass
parameter of the theory i.e. {\it $R$ is not an arbitrary parameter
to be introduced  ad hoc}.
 The important point
to note, from Martin's work, is that in the t-plane, in the intersection of
the regions of SLE and LLE, there is a circle of radius $R$ which is
independent of $s$. Thus the analyticity domain of $F^X$ is contained in this
region. Therefore, the amplitude is analytic in the domain:
$|t|<R \otimes~{\rm cut}~s$-plane.
\\   
Martin proved the theorem for $s$ below the threshold from the
BEG \cite{beg1,beg2} results. The analytic continuation is applicable for 
$0<s,s_{thr}=4m^2$. In order to proceed further, so that we can 
investigate the analyticity properties (and hence the growth properties)
of the amplitude, it is necessary to get information about the amplitude
beyond this rage of $s$. Sommer \cite{sommer1} provided the resolution. He
removed the cut $4m^2<s<s_1$ by defining a function  
\be
F^X(s,t)=F(s,t)- {1\over{\pi}}\int_{s_{thr}}^Xds'{{A_s(s',t)}\over{s'-s}}
\ee
in the range $-T\le t\le 0$. Now $F^X(s,t)$ has a right hand cut starting 
at $s=X$ (earlier the cut of $F(s,t)$ started from $s_{thr}=4m^2$). 
The positivity property of $F^X(s,t)$, so crucial to us, remains the same as
that of $F(s,t)$. Let us consider $s$ in the following region:
$4m^2<s_1<X$.
Notice that $F^X(s_1,t)$ is analytic in the domain which lies in the
intersection of the two functions
\be
F(s_1,t)~~{\rm and}~~
-{1\over{\pi}}\int_{s_{thr}}^Xds'{{A_s(s',t)}\over{s'-s}}
\ee
Thus $F(s_1,t)$ is analytic in the Lehamann region given by (\ref{martin24}).
Moreover, for complex $s$ 
\be
-{1\over{\pi}}\int_{s_{thr}}^Xds'{{A_s(s',t)}\over{s'-s}}
\ee
is analytic in the large Lehmann ellipse associated with 
$4m^2<s'<X$. We also know that, in the $cos\theta$-plane
 the semimajor axis of large Lehmann ellipse
shrinks with energy. What is important for us is that 
\be 
-{1\over{\pi}}\int_{s_{thr}}^Xds'{{A_s(s',t)}\over{s'-s}}
\ee
has still the same analyticity domain in $t$ \cite{sommer1} and it is a
distribution in $s$. Moreover, as has been proved \cite{sommer1,book2}
for the case of $D=4$, the domain of convergence in the $t$-plane is 
circle of radius $R$ and it is $s$-independent. Sommer's arguements are also
valid for the amplitude of D-dimensional theory i.e. it is analytic
inside a circle in the $t$-plane and the radius is $s$-independent.

\bigskip

\noindent{\bf 4.2 The Partial Wave Expansion and Asymptotic Behavior of 
Amplitude}

\bigskip
\noindent  We present the partial wave expansion for the scattering amplitude
in this subsection. Soldate \cite{soldate} 
considered graviton-graviton scattering in
arbitrary dimensions without accounting for their spins. He noted that
the amplitude admits a partial wave expansion with the Gegebauer polynomial as
the basis function. This function is also known as ultraspherical Jacobi
polynomial and its domain of convergence lies in the interval
$-1\le cos\theta\le +1$. We have argued earlier that for elastic scattering
of equal mass spinless particle there are only two admissible kinematic
variables (also holds for elastic scattering of 
unequal mass particles). The easiest way to arrive at
this conclusion is to note that the momenta of the 
four particles can be chosen to
lie is a four dimensional subspace of the 
D-dimensional space without loss of generality
as was argued in the preceding section (see remark (iv) after (\ref{edge})).
Once we impose mass shell conditions on scattering particles we are left with  only two
Lorentz invariant variables $s$ and $t$ which correspond to 
the c.m. energy squared and the
momentum transfer squared. If we consider, say, the case of $D=10$, then we
have to consider the rotation group $SO(9)$, ( in D-dimensions it is
$SO(D-1)$). The general cases have been studied in mathematical literature
\cite{lnr}. These authors have
considered the problem of representations in a series of paper. For
$9$-dimensional spherical harmonics, it will depend on angular momenta denoted
by $l_2,...l_5$, magnetic quantum numbers $m_1,...m_4$, correspondingly
there will be the  angles $\theta_2,...\theta_5$ and 
four azimuthal angles $\phi_1,...\phi_4$.
The 'spherical harmonics' will be given by
\be
\label{roton1}
Y^{l_2..l_5}_{m_1...m_4}(\theta_2,..\theta_5:\phi_1,...\phi_4)
\ee
We know that there is only one scattering angle and therefore, all other angles
can be integrated out. We are left with the Gegenbauer polynomial
denoted by: $ C^{\lambda}_l(cos\theta)$
The partial wave expansion is \cite{soldate}
\be
\label{roton2}
F^{\lambda}=A_1s^{-\lambda+1/2}\sum_{l=0}^{\infty}(l+\lambda)
f_l^{\lambda}C^{\lambda}_l(cos\theta)
\ee
The amplitude $F^{\lambda}(s,t)=F(s,t)$. Thus with this identification,
 the analyticity properties of  $F^{\lambda}(s,t)$ have been studied 
in the previous sections. 
 We introduce the index $\lambda$, where
$\lambda={1\over 2}(D-3)$, to keep track of the spacetime dimension we are
dealing with. Note that for $D=4$, $\lambda={1\over 2}$ and in this case the
Gegenbauer polynomial is the  Legendre polynomial.
$A_1= 2^{4\lambda+3}\pi^{\lambda}\Gamma(\lambda)$, independent of $s$ and $t$.
The factor $s^{-\lambda+1/2}$ on right hand side of (\ref{roton2}) begs
explanation. On the dimensional ground, we need this factor if we want
the partial wave amplitudes $\{f^{\lambda}_l \}$ to be dimensionless
in order to facilitate
the partial wave unitarity relation in the conventional form
\be
\label{roton3}
0\le |f^{\lambda}|^2\le~ {\rm Im}f^{\lambda}\le 1
\ee
 Moreover, eventually, when one derives the bound
\cite{moshe1,moshe2} on $\sigma_t$,
 the factor $s^{-\lambda+1/2}$ disappears in the expression for the bound.
$C_l^{\lambda}(x)$ are the Gegenbauer polynomials satisfying orthogonality
conditions with weight factor $(1-x^2)^{{\lambda}-1/2}$, $-1\le x\le+1$
\cite{szego}. The partial wave unitarity (\ref{roton3}) can be derived
through the standard procedure; using the orthogonality relations
of the Gegenbauer polynomials. Let us discuss, the positivity properties of
the absorptive amplitude. Recall that \cite{bateman1}(see p184)
\be
\label{roton4}
C^{\lambda}_l(1)={{\Gamma(l+2\lambda)}\over{{l!\Gamma(2\lambda)}}}
\ee
Again from \cite{bateman2} (see p206)
\be
\label{roton5}
{\rm Max}_{-1\le x\le +1}|C^{\lambda}_l(x)|=C^{\lambda}_l(1)
\ee
$C^{\lambda}_l(1)$ is positive,  therefore,
\bea
\label{roton6}
|{\rm Im}~F^{\lambda}(s,t)|=&& A_1s^{-\lambda+1/2}|
\sum_{l=0}^{\infty}(\lambda+l)
{\rm Im}f^{\lambda}_lC^{\lambda}_l(x)|\nonumber\\&&
\le A_1s^{-\lambda+1/2}\sum_{l=0}^{\infty}(\lambda+l){\rm Im}f^{\lambda}_l
|C^{\lambda}_l(x)|\nonumber\\&&
\le A_1s^{-\lambda+1/2}\sum_{l=0}^{\infty}(\lambda+l){\rm Im}f^{\lambda}_l
C^{\lambda}_l(1)\nonumber\\&&
=A_s^{\lambda}(s,t=0)
\eea
Here the notations are $x=cos\theta$ and
$A_s^{\lambda}(s,t)=A_s(s,t)$ as defined earlier and
$0\le{\rm Im}f^{\lambda}_l \le 1$. Thus we have proved the first inequality
of positivity for $A_s(s,t=0)\ge | A_s(s,t)|$ as promised. Now we need to show
${{d\over{{dt}}}}A_s(s,t)|_{t=0}\ge |{{d\over{{dt}}}}A_s(s,t)|$. This is
achieved from the derivative relation of the Gegenbauer polynomial
\cite{bateman2} 
\be
\label{roton7}
{{d\over{dx}}}C^{\lambda}_l(x)=2{\lambda}C^{\lambda+1}_{l-1}(x),
\ee
The above relation and eq. (\ref{roton5})  is  used to prove
${{d\over{{dt}}}}A_s(s,t)|_{t=0}\ge |{{d\over{{dt}}}}A_s(s,t)|$. For the
$n^{th}$ $t$-derivative  of the absorptive part,  we may
use the chain of relation (the first one) in the above equation repeatedly as
follows. First start with the expression for ${{d}\over{dt}}A^{\lambda}(s,t)$ 
which will involve first derivative of the Gegenbauer polynomial and use
(\ref{roton7}) to convert derivative of $C^{\lambda}_l$ to another
Gegenbauer polynomial. Then take the $t$-derivative of this 
expression to get second derivative of $A_s(s,t)$. We would derive the 
positivity property of
second $t$-derivative of $A_s(s,t)$. We then continue 
this chain of arguments to 
derive positivity property of the $n^{th}$ $t$-derivative of the absorptive
amplitude. 
   Thus, the postivity properties of the absorptive amplitudes 
that were  so useful to prove power series expansion in case of $D=4$ theories,
are also valid in arbitrary dimension, $D>4$. Therefore, the generalized
Martin theorem goes through as was shown already in this section.\\
Now we shall very briefly recapitulate the derivation of 
Froissart-Martin bound for
D-dimensional theories \cite{moshe1,moshe2}. We would like to remind that
the earlier result was derived under two assumptions:
 (AI) polynomial boundedness of the scattering amplitude and (AII) convergence
of partial wave amplitude inside an extended ellipse with semimajor axis
$1+{{2{\tilde T}_0}\over s }$, we use notation ${\tilde T}_0$ to distinguish 
it  from $T$ we have
introduced already. These assumption crucially used in \cite{moshe1,moshe2}. 
Here, they  have been derived 
 {\it ab initio}.\\
In the present investigation, we adopted LSZ formalism to discuss scattering
of massive, spinless particle in D-dimensions. In this approach, the amplitude
is a tempered distribution. Moreover, within the frame work of LSZ approach,
the Fourier transformed amplitude is polynomially bounded in momentum 
variables.  Moreover, we know how to
write down dispersion relation for such a case i.e. we might have to write
subtracted dispersion relations.\\
 We have shown the existence
of SLE and LLE within the LSZ frame work. Moreover, since we are dealing
with only a single type of particle (these are their antiparticles too) the
direct channel and the two crossed channels are the same. In such a crossing
symmetric theory, we have argued, following Martin, that the the radius of the
circle in the $t$-plane, $R$, is $4m^2$. To remind, the scattering amplitude
$F(s,t)$ is analytic in $t$  in quasi topological product
$\{|t|<R=4m^2 \}\otimes$ cut $s$-plane. Thus we have provided proof
of the two assumptions used in \cite{moshe1,moshe2} in the present
investigation. \\
We have not determined how many subtractions are required in the dispersion
relation i.e. what is the integer $N$ that appears in the dispersion relation.
In other words, can we write an unsubtracted dispersion relation or we need
to write a subtracted dispersion relation?
If the answer to second question is in affirmative,
the next question is how many subtractions we need?\\
In order to answer this question, we are required to determine the asymptotic
growth properties of the scattering amplitude, especially in the forward
direction. We begin by utilizing the property of polynomial boundedness of
the scattering amplitude.

\bigskip

\noindent{\bf 4.3 The High Energy Behavior of Scattering Amplitude}

\bigskip
\noindent We investigate the behavior of scattering amplitude at asymptotic
energies which is based on the results we have derived until now. (i)
The scattering amplitude is polynomially bounded in $s$ in the sense
that the dispersion integral is written with $N$ subtractions. We may take
$N$ to be even without loss of generality.
 (ii) The analyticity property of
the amplitude in the domain  $|t|<R$.  We recall the results,
in the context of Froissart-Martin bound \cite{moshe1,moshe2},
which were derived
with (i) and (ii) as {\it ad hoc} 
assumptions; presently (i) and (ii)  are not so.
Now on we shall take $R=4m^2$. Therefore, in the region $t={\bar R}$ with
${\bar R}=R-\epsilon$ we can write a dispersion relation since
\be
\label{roton8}
A_s^{\lambda}(s, t={\bar R})<{\bar C}s^N,~~~ {\bar C}= {\rm Constant}
\ee
Note that $A_s^{\lambda}(s,t)=A_s(s,t)$, we use this definition since there will
be some $\lambda$-dependent constants i.e. D-dependent constants as we proceed.
We recall that $A_s^{\lambda}(s, t={\bar R})$ is analytic in this region in
$t$-plane. Thus the partial wave expansion
\be
\label{roton9}
A_s^{\lambda}(s,t={\bar R})=A_1s^{-\lambda+1/2}\sum_{l=0}^{\infty}(l+\lambda)
{\rm Im}~f^{\lambda}_ls)C_l^{\lambda}(1+{{\bar R}\over{{2K^2}}})
\ee
converges since  $A_s^{\lambda}(s, t={\bar R})$ is analytic in the domain
$|t|<R\otimes$cut $s$-plane. We also know the large-$s$ behavior of 
 $A_s^{\lambda}(s, t={\bar R})$ inside Lehmann ellipse. Note that the over all factor
 ${{\sqrt s}\over K}$ which usually appears in definition of the amplitude
has been dropped since this ratio is
equal to $1$ in the large $s$ limit. Moreover, in the forward direction
\be
\label{roton10}
A_s^{\lambda}(s,t=0)=A_1s^{-\lambda+1/2}\sum_{l=0}^{\infty}(l+\lambda)
{\rm Im}~f^{\lambda}_ls)C_l^{\lambda}(1)
\ee
There is a constant  positive factor  $C_l^{\lambda}(1)$ appearing in the 
above equation.  This is the starting point to prove the Froissart bound which Martin
improved by exactly determining certain constants. We shall not go through
 all the steps since this has been undertaken by \cite{moshe1,moshe2}.
They adopt the same maximization program proposed by Martin \cite{book2};
 however, for
general D-dimensional case there are departures which we shall point out
in sequel.
The extremization is achieved by resorting to Martin's method.
(i) Choose ${\rm Im}~f^{\lambda}_l(s)=1$ for $0\le l\le L$ (ii)
${\rm Im}~f^{\lambda}_l(s)=\epsilon<1$ for $l=L+1$. (iii) And
${\rm Im}~f^{\lambda}_ls)=0$ for $\l>L+2$. Here we consider the case when
the ratio ${L\over{\sqrt s}} \rightarrow \infty$ for large $s$ which
eventually leads to Froissart-like bound i.e. the total cross section
is bounded by power of $ln s$. The other situation where the ratio 
${L\over{\sqrt s}}$ goes to a constant would make total cross section
bound by a constant. As is well known, the polynomial
boundedness and the partial wave expansion (\ref{roton10}) are crucial
ingredients to choose the cut-off value of $L$. In contrast to $D=4$ case, where
one dealt with Legendre polynomials, there are some departures to determine
the cut off, $L$. The large $l$ behavior of
$C_l^{\lambda}(1+{{\bar R}\over s})$ is
\be
\label{roton11}
 C_l^{\lambda}(1+{{\bar R}\over s})\sim e^{2l\sqrt{{\bar R}/s}}l^{\lambda -1}
({{s\over{R}}})^{{{\lambda}\over 2}} G(\lambda)
\ee
where $G(\lambda)$ is a function which depends only on $\lambda$; we shall
display it whenever necessary. Noting the polynomial boundedness property
(\ref{roton8}) and the large $l$ behavior of the Gegengauer polynomials for 
the argument  greater than $1$, we get
\be
\label{toton12}
A_s^{\lambda}=2^{4\lambda+3}\pi^{\lambda}\Gamma(2\lambda)
s^{-{\lambda\over2}+1}({\bar R})^{-{{\lambda+1}\over 2}}
e^{2L{\sqrt{{R\over s}}}}L^{\lambda}\le {\bar C}s^N
\ee
Thus we find that the cut-off value, $L$, is
\be
\label{roton12}
L={1\over 2}{\sqrt{{s\over{{\bar R}}}}}lns + ~{\rm ~terms~ nonleading~ in}~
lns
\ee
A remarkable feature is the energy dependence of the cut-off
$L\sim {\sqrt s}lns$  \cite{moshe1,moshe2}; there is no power of $\lambda$
in energy dependence.  This $s$-dependence is the  same as
in the $4$-dimensional theory.  The bound on $A_s^{\lambda}(s,t=0)$ now
follows \cite{moshe1,moshe2}
\be
\label{roton13}
A_s^{\lambda}(s,t=0)=\sum_0^L(l+\lambda)C^{\lambda}_l(1)\le B(\lambda)
\Psi(N,R)s(lns)^{D-2}
\ee
where  ${\bar R}=4m^2-\epsilon$.
\be
\label{roton14}
B(\lambda)={{2^{\lambda}\Gamma(\lambda)\Gamma(\lambda+1/2)}\over{{
\pi^{3/2}\Gamma^2(2\lambda)}}},~~
\Psi(N,{\bar R})=\bigg({1\over 2}{{(N-1)}\over{{\sqrt{{\bar R}}}}} \bigg)^{D-2}
\ee
$\lambda={1\over 2}(D-3)$.
The total cross section is bounded from the above as 
\be
\label{roton14}
\sigma_{\rm total}\le B(\lambda)\Psi(N,{\bar R})(lns)^{D-2}
\ee
One interesting feature of the bound is its energy dependence i.e. is a
power of $ln s$. Thus for the four dimensional case one recovers the
 the Froissart-Martin
bound, $(lns)^2$. It is worth while to mention
that the upper bound (\ref{roton14}) contains an  unknown
parameter. I do not consider $B(\lambda)$ as an unknown function since it gets
fixed once we decide the dimensionality of spacetime we work in. However,
$\Psi$ depends on $N$;
the number of necessary subtraction is not determined so far.\\
In order to derive what  value $N$ takes, let us consider the modulus of the
forward scattering amplitude and expand it in partial waves. 
First of all, we can cut off the partial wave sum at $L$. Thus
\be
\label{roton15}
|F^{\lambda}(s,t=0)|\le{\large\sum}_0^{L}(l+\lambda)C^{\lambda}_l(cos\theta=1)
|f^{\lambda}_l(s)| +{\rm terms~ with~sum~ starting}~ l>L+1
\ee
where $L={{\sqrt{s\over{{\bar R}}}}} {1\over 2}(N-1) ln s$. Thus the remainder
of the sum starting $L+1$ can be made as small as we desire.
It is understood that the right hand side of 
the above equation might have constant
prefactors; however, their presence will not affect the ensuing discussions.
Using partial wave inequality, we conclude from (\ref{roton13}) that
\be
\label{roton16}
|F^{\lambda}(s,t=0)|< {\rm Constant}~s(lns)^{D-2}
\ee
Remark:  This bound is generalization of Jin and Martin
\cite{jinmart} bound to D-dimensions.
As we have argued elsewhere, crossing symmetry is valid for the case under
study and invoking crossing, we conclude that the modulus of the forward
scattering amplitude $|F(s,t=0)|< |s|(lns)^{D-2}$. The bound holds 
on the right hand cut as well as  on the left hand cut. Thus
$F^{\lambda}(s,t=0)$ is polynomially bounded in the complex $s$-plane. Now
invoke Phragman-Lindelof theorem \cite{tit}: $|F^{\lambda}(s,t=0)|$
 is bounded by
${\rm Constant}~s(lns)^{D-2}$ in the entire complex $s$-plane. Therefore,
we need at most two subtractions, i.e. $N=2$, not only in the forward
direction, $t=0$ but for $-T\le t\le0$. Moreover, for $|t|<R$ the number of
subtractions, $N=2$ (even) is conserved, this is true in the complex
$s$-plane. We have now fixed $N=2$. Therefore, our work can be summarized as :\\
{\it Theorem:} For a massive neutral scalar field theory 
which satisfies axioms of
Lehmann, Symanzik and Zimmermann formalism, 
the upper bound on the total cross section, $\sigma_t$,
is
\be
\label{roton17}
\sigma_{\rm total} \le B(\lambda)
\bigg({{1\over{{2\sqrt{4m^2-\epsilon}}}}}\bigg)^{D-2}(lns)^{D-2}
\ee
where $B(\lambda)={{2^{\lambda}\Gamma(\lambda)\Gamma(\lambda+1/2)}\over{{
\pi^{3/2}\Gamma^2(2\lambda)}}}$.

\bigskip

\noindent {\bf 5. Summary and Discussions }
\bigskip

\noindent We summarize our results and discuss the consequences.  We began with an
intent to derive the high energy behavior of scattering amplitude in
D-dimensional massive field theories. Our principal goal was to remove certain
arbitrariness in the derivation of the bound on total cross section which
was obtained earlier \cite{moshe1,moshe2}. Essentially, there were two
assumptions which were not proven in the field theoretic 
frame work. We have proven
that these two assumptions can be derived from LSZ formalism. In order to
arrive at our goal, we needed the  edge-of-the-wedge theorem. We have argued
that the theorem is likely to hold so long as we consider the four point
scattering amplitude. 
We have argued that  the proof of Bremermann, Oehme and Taylor \cite{bmt}
will  also be 
valid in D-dimensional theories. We have presented the supportive 
arguments in the preceding section.
 In fact, by choosing suitable coordinate frame, in case of
four particle amplitude, we can confine to a four dimensional 
subspace of the D-dimensional
momentum space. Subsequently, the BEG wedge-of-the-edge theorem will be
proved as has been argued in Section 3. The second important result 
used by us is the analog
of BEG \cite{beg2} theorem. We have argued 
 regarding existence of analyticity domain in the 
neighborhood of $s$ and $t$ just below $s_{thr}$ to prove Martin's theorem. 
Once
again, in case of four point amplitude, 
if we confine ourselves to a $4$-dimensional momentum subspace,
 as alluded to
above, this theorem will also be valid. 
We have not presented explicit proofs of these two 
results for the D-dimensional theory.
However, we feel that 
the arguments are adequate to utilize the results of these theorems for
our purpose. It is worth while to point out that we adopted the LSZ formulation 
to achieve
this goal without resorting to any specific model. 
It is assumed that there are no
bound states in this theory.  \\
The strategy adopted to derive the asymptotic behavior 
of scattering amplitude is
as follows. As a first step, it was  necessary to establish that a fixed-t
dispersion relation can be written for the scattering amplitude. In order to
reach this goal, the  essential step was 
to prove that the absorptive part of
the amplitude is well behaved for fixed physical $t$ as
$s\rightarrow s_{thr}$. We showed that there are ellipses in the $t$-plane
where the amplitudes have desired behavior. In particular, starting from
the LSZ reduction technique we showed the existence of a large ellipse in
D-dimensional theory which is analogous to the large Lehmann ellipse.
In order to prove the existence of the ellipses, we needed to prove the
existence of Jost-Lehmann-Dyson representation for the retarded function.
We proved the generalized Dyson theorem to achieve our goal.\\
We have accomplished the target of establishing the dispersion relations
in $s$ for fixed $t$. We needed to prove a generalized version of Martin's
theorem to derive constraints on the growth properties of scattering
amplitude as a function of $s$. It was shown that, indeed there is a circle
  inside the domain of analyticity in the $t$-plane inside which the scattering
amplitude, $F^{\lambda} (s,t)$, 
can be expanded in a power series in $t$ and the power
series converges absolutely. We also proved positivity properties of
the absorptive part of the amplitude and its $t$-derivatives for the
D-dimensional case by exploiting some of the properties of the Gegenbauer
polynomials. 
This is achieved, after we expanded the amplitude in the
basis of the Gegenbauer polynomial. 
 Thus the generalized version of Martin's theorem
could be proved for the D-dimensional field theories.\\
The asymptotic growth properties of the amplitude had been investigated
in \cite{moshe1,moshe2} under the assumptions (AI) and (AII) as stated
in Section 1.  These 
assumptions played central role in derivation of the bounds in
\cite{moshe1,moshe2}. We recall that these authors had assumed  the
existence of an 'analog' Lehmann ellipse whose semimajor axis is characterized
by a constant ${\tilde T}_0$ which is independent of $s$. In the present work
we have proved existence of such a domain of analyticity 
i.e. the Large Lehmann Ellipse
(LLE). We may remind that
such a parameter, $t_0$, also appears in four
dimensional theories; however, it is determined from the 
first principles. In most
of the hadronic processes it turns out to be $4m_{\pi}^2$ \cite{am}.
Moreover, Sommer \cite{sommer1}
has given a prescription to determine $t_0$. The second assumption   
\cite{moshe1,moshe2} is the polynomial boundedness of the scattering amplitude,
$|F^{\lambda}(s,t)|<s^N$ inside a certain ellipse \cite{moshe1,moshe2}.
We have proved, within LSZ axioms, that  the amplitude is polynominally
bounded (due to the temperedness) and the number of required subtractions is
$N=2$.  In nutshell, the
work reported in  \cite{moshe1,moshe2} left two important questions to be
answered: (I) what is value of $R$?, in our notation and (II) what is
value of N?\\
Our long investigation has provided definite answers to these questions as
was presented in Section 4. We showed that $R=4m^2$.
 Moreover, the value is determined from
LSZ formalism together with Martin's analysis. We demonstrated that $N=2$.
Again, having proved the asymptotic growth properties of the absorptive
amplitude in $s$ in a domain $|t|<R$, one can show how the forward scattering
amplitude is bounded i.e. the asymptotic 
behavior of $F^{\lambda}(s,t=0)$. Furthermore, from
the bound on $A_s^{\lambda}(s,|t|<R)$, we can impose a constraint
on the scattering amplitude in the same $t$ domain. Finally, as we have shown,
the scattering amplitude needs at most two subtractions i.e. $N=2$. Therefore,
the bound  we have derived now  
has no free parameters. This statement is to be made
with a qualifying remark that there is the unknown energy scale which
 is necessary
to scale $s$ i.e. $(lns)^{D-2}\rightarrow [ln({s\over{s_1}})]^{D-2}$ 
in the bound which is not
fixed from first principles.
Note that corresponding scale for four dimensional theories is also not
determined from first principles of quantum field theory. We feel that it is
quite satisfying that both the unknown parameters are now determined in
the frame work LSZ formalism.\\
Furthermore, the upper bound \cite{jmjmp} on $|F^{\lambda}(s,t)|$ for $|t|<R$,
which was  deduced for large $s$ is 
now established from the results proved here.
No {\it additional} assumption is required. It is also important to mention
that the upper bound and lower bound derived by me for the 
absorptive amplitude (the second theorem)
\cite{jmjmp} now needs no extra assumption.  Indeed, the results derived in the
 present 
work removes what was termed as an extra assumption for derivation
of the two bounds for the absorptive 
amplitude in a small $|t|$ region including the physical domain.
Now the theorems of \cite{jmjmp} can be utilized to derive new bounds on
elastic differential cross sections and  bounds on slope of diffraction peak.
Moreover, since the differential cross sections have been measured at the
LHC enegies the scaling behavior of differential cross sections might be
explored.  \\
 We recognize that it might be possible to derive the results presented
here through more formal approach to axiomatic field theories. 
In such frame works
some of the assumptions such as the field 
operators being operator values distributions
 are not invoked. In other 
words the temperedness property of the amplitude is not
required in some of these formulations. It might be possible to prove 
the polynomial boundedness of the scattering
amplitude as derived by Epstein, Glaser and Martin \cite{egm}.\\ Now we present 
some arguments and the phenomenological scenario in the context of the present
investigation.
Let us envisage the scenario of low scale compactification. In this
proposal, the scale of compactification could be as low as $500$ GeV or
$1$ TeV. In other words, the extra dimensions decompactify at this energy
scale. Consequently, the decompactification effects could manifest in very
high energy accelerator 
experiments. The lowest mass particles will have mass-value same
as this scale. Therefore, one could argue that $s_1 \sim s_{comp}$. This
is a plausible proposal. We may ask: if the decompactification scale is so low
can we get some hints of this low scale? The phenomenology of this scenario
has already been worked out in some details \cite{anto,luest}. 
We had proposed another
scenario to experimentally investigate existence of low scale compactification
proposals.
Consider high energy collisions in an energy scale above decompactification
scale. Then there is a possibility that effect of higher dimensions, $D>4$,
 might
manifest in high energy scatterings. In particular such an effect might show
certain high energy behaviors unfamiliar to us. For example the data for
total cross sections might seemingly 
violate the Froissart bound derived for $D=4$
theories i.e. $\sigma_t\sim (lns)^2$; in fact the data fits this behaviors
over wide energy range i.e.  $\sigma_t$ exhibits a $(lns)^2$ 
behavior in very high energy processes.
 In order to explore the possible signal for
decompactification at low energy scale ( $500$ GeV to $1$ TeV range) one should
examine the energy dependence of total cross sections in collision energies
above ${\sqrt s}> 500$ GeV and try to fit
 with a phenomenological formula for $\sigma_t$. 
Thus $\sigma_t$ might assume  a form  \cite{jmjmp1}
\be
\label{fitbound}
\sigma_t=\sigma_0+ C_1(ln{s\over{s_0}})^2+ C_2(ln{s\over{s_1}})^{\beta}, ~~
 \beta>2
\ee
Here $\sigma_0$ corresponds to a constant, independent of $s$, the so called
Pomeranchuk term. The next term is the Froissart bound-like
energy dependence.  
The last term encodes effect of higher dimension decompactification.
 If we find a fit with such a parametrization, it might provide an
indirect evidence for low scale compactification. It is fair to take
$s_0$ in the range of decompactification scale. We tried to obtain a
qualitative fit to total cross sections of old LHC data at $6$ TeV and $7$ TeV
together with very high energy cosmic ray data. We found \cite{jmjmp1} that with
$\beta \sim 2.3$ we get  a fit with reasonable $\chi^2$. 
However, the cosmic ray data are reported
with large error bars and therefore, the available set of data  to fit
cross sections is not large enough to conclude  that $\beta>2$. 
  We feel that a more careful procedure to  fit 
 very high energy
experimental data with the forthcoming results from the LHC 
might be a promising endeavor to explore the hypothesis
of low scale decompactification at accelerator energies.\\
Indeed, it will be quite interesting 
to study analyticity properties of scattering 
amplitude in a higher dimensional 
theory where some of the spatial coordinates are
compactified. It was pointed out, in the context of potential scattering 
\cite{khuri1,khuri2}, that
for nonrelativistic theory, the analyticity properties of scattering
amplitudes are different from those of a nonrelativistic theory which has no 
compactified spatial coordinates. It is to be noted that momenta associated with
compact directions are discrete. Moreover, the deviation from usual dispersion
relations for certain potential models (with compact coordinates) raised the
question that for field theories with compactified coordinates might not satisfy
the known dispersion relations \cite{khuri1}. 
It was argued that these effects might be observed
in high energy scatterings at LHC energies. 
Therefore, it is worth while to pursue
these issues for a $D$-dimensional field theory where certain 
spatial coordinates are
compact. The present investigation can be utilized towards this end.\\
\noindent 
{\bf Acknowledgments}: I am grateful to Henri Epstein for very valuable
discussions and for sharing his insights.  I would like to thank Andr\'e 
Martin for discussions. I have benefited from discussions with Chand Devchand,
Hermann Nicolai and Stefan Theisen at various stages of this project.
The work was initiated at the Max-Planck Institute of Physics ( Werner 
Heisenberg Institute), Munich during a short visit. I thank the Institute
and Deter L\"ust for their hospitality. Most of the research was carried out
at the Max-Planck  Institute of Gravitational Physics ( Albert Einstein
Institute),  Golm. I thank Hermann Nicolai and the members of the Institute
for their very gracious hospitality. This work is partially supported by
the Indian National Science Academy, New Delhi, through the Senior Scientist
Programme.

\newpage
\begin{center}
{\bf Appendixes}
\end{center}
\bigskip

\noindent {\bf Appendix A: Proof of Dyson's Theorem: Generalized to 
D-dimensions}

\bigskip
\noindent We have derived the Jost-Lehmann representation for causal
function $F_C(q)$ and for $F_R(q)$ in the D-dimensional theory. This was
achieved in the LSZ formulation. Moreover, we analyzed the location of the
singularities of $F_R(q)$ generalizing the approach of Jost and Lehmann.
It is worth mentioning that Jost-Lehmann representation is valid for the case
of equal mass particle.\\
Dyson \cite{dyson} used an indigenous technique to derive the representation
for the case of unequal mass in a more elegant mathematical frame work. We
have generalized Dyson's formalism for theories in arbitrary dimensions 
which satisfy LSZ axioms. We mention in passing that
Dyson's derivation was also based on  the LSZ formulation of field theories.
 To recapitulate,
we have derived the expressions for $F_R(q), F_A(q) ~{\rm and}~F_C(q)$ already
for the case of D-dimensional theories in Section 3. 
Furthermore, the support properties
of these functions in their Fourier transformed coordinate space have been
alluded to in that section. We consider D-dimensional Lorentzian space 
time manifold
and supplement it with 
two extra spatial signature coordinates in order to  generalize Dyson's
formalism. The coordinates of the $(D+2)$-dimensional spacetime are
\be
\label{dyson1}
{\tilde z}=\{{\tilde z}_0=x_0,{\tilde z}_1=x_1...{\tilde z}_{D-1}=x_{D-1},
{\tilde z}_D=y_1, {\tilde z}_{D+1}=y_2 \}
\ee
The $(D+2)$-dimensional momenta are defined as
\be
\label{dyson2}
{\tilde r}=\{{\tilde r}_0=q_0,{\tilde r}_1=q_1,...{\tilde r}_{D-1}=q_{D-1},
{\tilde r}_D=p_1, {\tilde r}_{D+1}=p_2 \}
\ee
The metric is: ${\rm diag}~(+1,-1,...-1)$ and
\be
\label{dyson3}
{\tilde z}^2={\tilde x}^2-y^2=x_0^2-x_1^2-...-x_{D-1}^2-y_1^2-y_2^2
\ee
We recall ${\tilde F}_C(x)$ is the Fourier transform of $F_C(q)$. Now we
define ${\tilde F}_C({\tilde z})$ in $(D+2)$-dimensions from the given
D-dimensional function ${\tilde F}_C(x)$
\bea
\label{dyson4}
{\tilde F}_C({\tilde z})=&&4\pi{\tilde F}_C(x)\delta(x^2-y^2)\nonumber\\&&
=4\pi{\tilde F}_C(x)\delta({\tilde z}^2)
\eea
We note that ${\tilde F}_C({\tilde z})$ is defined on the light cone of the
$(D+2)$-dimensional ${\tilde z}$-space.
\bea
\label{dyson5}
\int_{-\infty}^{+\infty}\int_{-\infty}^{+\infty} dy_1dy_2{\tilde F}_C({\tilde z})&&
=4\pi^2{\tilde F}_C(x)~, {\rm for}~x^2\ge 0 \nonumber\\
&& =   0,~{\rm for}~x^2<0
\eea
We have constructed ${\tilde F}_C({\tilde z})$ in $(D+2)$-dimensions. Notice
that by construction ${\tilde F}_C({\tilde z})$ and ${\tilde F}_C(x)$
(  ${\tilde F}_C(x)=0,~{\rm for}~x^2<0$) are equivalent in the sense that
we may recover  ${\tilde F}_C({x})$ from ${\tilde F}_C({\tilde z})$
by integrating over $d^2y$ (see eq. (\ref{dyson5})). Now we choose a special
$(D+2)$-dimensional momentum vector:
\be
\label{dyson6}
{\hat q}=(q_0,q_1,...q_{D-1}, 0, 0)
\ee
We have set last two components of $\tilde r$ to zero with this choice. The
Fourier transform of ${\tilde F}_C({\tilde z})$, defined as
${\bar F}_C({\tilde r})$,
 is given by
\be
\label{dyson7}
{\bar F}_C({\tilde r})={1\over{(2\pi)^{D+2}}}\int e^{i{\tilde r}.{\tilde z}}
{\tilde F}_C({\tilde z}) d^{D+2}{\tilde z}
\ee
Let us insert expression for  ${\tilde F}_C({\tilde z})$, (\ref{dyson4}), into
the Fourier transform(\ref{dyson7})
\bea
\label{dyson8}
{\bar F}_C({\tilde r})=&&{{4\pi}\over{(2\pi)^{D+2}}}\int  d^{D+2}{\tilde z}
d^Dqe^{i{\hat q}.{\tilde z}} {\bar F}_C(q)\nonumber\\&&
=\int D^{(1)}({\tilde r}-{\hat q}){\bar F}_C(q)d^Dq
\eea
where
\bea
\label{dyson9}
D^{(1)}({\tilde r})=&&{{2\over{(2\pi)^{D+1}}}}\int e^{-i{\tilde r}.{\tilde z}}
\delta({\tilde z}^2)d^{D+2}{\tilde z}\nonumber\\&&
={2\over{(2\pi)^{D+1}}}P~{{1\over{(\tilde r})^{D/2}}}
\eea
P stands for the principal value.
From now on, I shall drop the prefactors like
${{1\over{(2\pi)^{D+2}}}}, ~{{1\over{(2\pi)^{D+1}}}}$ etc. which come from
 taking Fourier transforms. As the next equation will show, we derive
 an expression for the ${\bar F}_C({\tilde r})$ which will display the
singularity structure and location of singularities in the $q$-plane. Now
use the expression for $D^{(1)}({\tilde r})$ in the above equation
\bea
\label{dyson10}
{\bar F}_C({\tilde r})=&&\int d^Dq{{F_C(q)}\over{\bigg(({\tilde r}-{\hat q})^2
\bigg)^{D/2}}}\nonumber\\&&
=\int d^Dq{{F_C(q)}\over{[(u-q)^2-{\bar s}]^{D/2}}}
\eea
where ${\bar s}=p_1^2+p_2^2$, ($\{p_1,p_2\}$ are
momenta  along extra
directions). It is important to remember that
${\tilde F}_C({\tilde z})= {\tilde F}_C(x)\delta({\tilde z}^2)$ whose support is
on the light cone of the $(D+2)$-dimensional spacetime. Moreover, 
the Fourier transformed
${\bar F}_C({\tilde r})$ is rotationally invariant on the
${\tilde r}_D-{\tilde r}_{D+1}$ plane since it depends on
${\bar s}=p_1^2+p_2^2$. A crucial observation,
originally due to Dyson \cite{dyson}, 
  is that
$D^{(1)}({\tilde r})$ satisfies a $(D+2)$-dimensional wave equation in the
momentum space
\be
\label{dyson11}
{{\large\Box}}_{D+2}D^{(1)}({\tilde r})=0,~~~{\rm where} ~~~{{\large \Box}}_{D+2}=
{{\partial^2}\over{{\partial}{{{\tilde r}_0}^2}}}-
\sum_{k=1}^{D+1}{{{\partial^2}\over{{\partial}{{{\tilde r}_k}^2}}}}
\ee
Furthermore, ${\bar F}_C({\tilde r})$ also satisfies the $(D+2)$-dimensional
wave equation: ${{\large\Box}}_{D+2}{\bar F}_C({\tilde r})=0$. The argument
of Dyson can invoked: if ${\tilde F}_C(x)$ vanishes for $x^2<0$, then
$F_C(q)$ is the boundary value of ${\bar F}_C(q)$ on ${\bar s}=0$ plane.
In other words, ${\bar F}_C({\hat q})=F_C(q),~{\hat q}=(q_0,...q_{D-1},0,0)$.
Moreover,
\bea
\label{dyson12}
{\bar F}_C({\hat q})=&&\int d^{D+2}{\tilde z}e^{i{\hat q}.{\tilde z}}
4\pi\delta(x^2-y^2){\tilde F}_C(x)\nonumber\\&&
=\int d^Dxe^{iq.x}4\pi\theta(x^2){\tilde F}_C(x)
\eea
This is achieved after integrating $\int dy_1dy_2$ and setting
${\hat q}.{\tilde z}=q.x$. Now
$F_C(q)=\int d^Dq'{{F_C(q')}\over{ [(q-q')^2]^{D/2}}}$. Recall that there is
a class of solution of $F_C(q)$ whose Fourier transform, ${\tilde F}_C(x)$,
 vanishes for $x^2<0$. Let this class be denoted by ${\cal C}$.
 Dyson's arguments
can be generalized for D-dimensional case as follows: a necessary condition
for ${\tilde F}_C(x)$ is that  it satisfies the micro causality property.  
${\bar F}_C(q)$ should be the boundary value on ${\bar s}=0$ plane of a
solution ${\bar F}_C(q,{\bar s})$, where  ${\bar F}_C(q,{\bar s})$
is a solution to the $(D+2)$-dimensional wave equation in the momentum
space. We have observed that this class of solutions has to be rotationally
symmetric in the plane ${\tilde r}_D-{\tilde r}_{D+1}$. Note that
${\bar s}=0$ is a time-like surface. Moreover, the boundary value of the
hyperbolic equation $\Box_{D+2}{\bar F}_C({\tilde r})=0$ is not arbitrary
on this surface. Alternatively,  a more general approach is to consider
a function which satisfies $(D+2)$-dimensional wave equation in the
${\tilde r}$-space and is rotationally invariant on the $(D)-(D+1)$ plane.
Its Fourier transform being 
\be
\label{dyson13}
{\tilde F}({\tilde z})=\int d^{D+2}{\tilde r}e^{-i{\tilde r}.{\tilde z}}
{\bar F}({\tilde r})
\ee
The Fourier transformed ${\tilde F}({\tilde z})$ is endowed with the
following features:  since  $\Box_{D+2}{\bar F}({\tilde r})=0$; therefore,
${\tilde F}({\tilde z})=\delta({\tilde z}^2)G({\tilde z}^)$ and it is note
worthy that ${\tilde F}({\tilde z})$ has its support on the light cone
of the $\tilde z$-spacetime.  Moreover, ${\bar F}({\tilde r})$ has
rotational symmetry on a plane as noted earlier. Therefore,
\bea
\label{dyson14}
{\tilde F}({\tilde z})=&&\int d^{D+2}{\tilde r}e^{-i{\tilde r}.{\tilde z}}
{\bar F}(u,|p|)\nonumber\\&&
=\int d^Du e^{-iu.x}\int_0^{\infty}pdp\int_0^{2\pi}e^{p|y|cos\theta}
{\bar F}(u,|p|)
\eea
in the polar decomposition of $(p_1,p_2)$ with ${\bar s}=p_1^2+p_2^2$.
Consequently,
\be
\label{dyson15}
{\tilde F}({\tilde z})=2\pi\int d^Du e^{-iu.x}\int_0^{\infty}d{\bar s}
J_0({\sqrt{{\bar s}}}|y|){\bar F}(u,{\bar s})
\ee
The Bessel function admits a power series expansion
$J_0({\sqrt{{\bar s}}}y)=\sum_0^{\infty}{{({\bar s}|y|)}\over{n!}}$; it
already shows the rotational invariance in the $y$-plane since it
depends on $y^2$. Moreover, from its structure
$ {\tilde F}({\tilde z})=\delta({\tilde z}^2)G(x,y^2)=
\delta(x^2-y^2)G(x,y^2)$;  
we  may conclude
\be
\label{dyson16}
 {\tilde F}({\tilde z})=\delta(x^2-y^2){\tilde f}(x)
\ee
With the above developments, ${\tilde f}(x)$ need not vanish for $x^2<0$.
If we want to relate it to causal function, we have to impose  the additional
condition: 
${\tilde f}(x)=0,~{\rm for}~ x^2<0$ from outside and then we  can identify
${\bar F}_C({\hat q})=F_C(q)$. Under this constraint we arrive at 
$ {\tilde F}_C({\tilde z})={\tilde F}_C({\tilde z})$. We can now present
the generalized version of Dyson's condition: the necessary and
sufficient condition  for a function to vanish outside the light come
of D-dimensional spacetime, i.e. $x^2<0$ is that $F_C(q)$ be the boundary
value on the surface ${\bar s}=0$ of a solution to a $(D+2)$-dimensional
wave equation ${\bf{\Box}}_{D+2}{\bar F}=0$. The solution is
required  to be rotationally invariant in $p_1$-$p_2$ plane.\\
In accordance with Dyson's prescription, in general, a solution to the
wave equation (of the type being discussed) can be expressed in terms of
its value and its normal derivative on an arbitrary spacelike surface. Thus,
for the D-dimensional case, we also introduce a singular function and
denote it as ${\bar D}({\tilde r})$. It also satisfies the homogeneous
wave equation
\be
\label{dyson17}
{\bf{\Box}}_{D+2}{\bar D}({\tilde r})=0
\ee
The initial conditions are
\be
\label{dyson18}
{\bar D}({\tilde r}_0=0,{\tilde r}_1,...,{\tilde r}_{D+1})=0,~~{\rm and}~~
{{{{\partial {\bar D}}({\tilde r})}\over{{\partial}{\tilde r}_0}}}\bigg|_{{\tilde r}_0=0}=
{\bf \Pi}_{i=1}^{D+1}\delta({\tilde r}_i)
\ee
We can write (\ref{dyson18}) explicitly as
\be
\label{dyson19}
{\bar D}({\tilde r})=\int d^{D+2}{\tilde z}e^{-i{\tilde r}.{\tilde z}}
\epsilon({\tilde z})\delta({\tilde z}^2)
\ee
Now we can choose a spacelike surface, $\Sigma$, and prescribe initial
data on it. If ${\bar F}({\tilde r})$ is solution to the wave equation.
Let it assume the value  ${\bar F}({\tilde r}')$ and
$\bigg({{\partial{\bar F}({\tilde r}')}\over{{\partial}{{\tilde r}'_{\alpha}}}}\bigg)n^{\alpha}({\tilde r}')$ on           
$\Sigma$ (where $n^{\alpha}$
is normal to the surface). Then
\be
\label{dyson20}
{\bar F}({\tilde r})=\int_{\Sigma}d\Sigma_{\alpha}\bigg[{
\bar F}({\tilde r}'),{{\partial\over{{\partial}{{\tilde r}'_{\alpha}}}}}
{\bar D}({\tilde r}'-{\tilde r}) \bigg]
\ee
We define
\be
\label{dyson21}
\bigg[{\bar F}({\tilde r}'), {{\partial\over{{\partial}{{\tilde r}'_{\alpha}}}}}
{\bar D}({\tilde r}'-{\tilde r}) \bigg]={\bar F}({\tilde r}')
{{\partial\over{{\partial}{{\tilde r}'_{\alpha}}}}}
{\bar D}({\tilde r}')- {{{\partial}{\bar F}({\tilde r}')}\over{
{\partial}{\tilde r}'_{\alpha}}}{\bar D}({\tilde r}')
\ee
Here $d\Sigma_{\alpha}$ is the surface element and it is a
$(D+2)$-dimensional vector normal to the spacelike surface.\\
We can derive the solution to the wave equation with assigned symmetry
properties by choosing the surface appropriately and with desired
boundary values for the solution (\ref{dyson21}). \\
Our original goal is to derive a representation for $F_C(q)$. Therefore,
we set ${\bar F}_C({\hat q})=F_C(q)$. The integral equation for the latter
is
\bea
\label{dyson22}
F_C(q)&&=\int_{\Sigma}d\Sigma'\bigg[{\bar F}({\tilde r}'),
{{\partial\over{{\partial}{{\tilde r}'_{\alpha}}}}}
{\bar D}({\tilde r}'-{\hat q}) \bigg]\nonumber\\&&
=\int_{\Sigma}d\Sigma'_{\alpha}\bigg[{\bar F}({\tilde r}'),
{{\partial}\over{{\partial}{{\tilde r}'_{\alpha}}}}\bigg\{\epsilon(u_0-q_0)
\delta'\bigg( (u-q)^2-{\bar s}\bigg)\bigg\}\bigg]
\eea
This representation is unique, as has been argued by Dyson. We have  a
function$F_C(q)$ (for which ${\bar F}_C(x) $ has the desired support
property) and we have a given surface $\Sigma$ and $F_C(q)$ admits the
representation (\ref{dyson22}) with any function
${\bar F}_C({\tilde r})={\bar F}_C(u,{\bar s})$ so that it depends on the
invariant ${\bar s}=p_1^2+p_2^2$. Moreover ${\bar F}_C({\tilde r})$
satisfies the desired wave equation in $\tilde r$-space then
${\bar F}_C({\tilde r})$ (is identical to ${\bar F}_C({\tilde r})$) defined by
\be
\label{dyson23}
{\bar F}({\tilde r})= \int d^Dq{{F_C(q)}\over{{[({\tilde r}-{\hat q}]^{D/2}}}} =\int {{d^DqF_C(q)}\over{[(u-q)^2-{\bar s}]^{D/2}}}
\ee
and
\be
\label{dyson24}
{\bar F}({\tilde r})=\int {\bar D}^{(1)}({\tilde r},{\hat q})
{\bar F}_C ({\hat q})d^Dq
\ee
I have suppressed the factors of $(2\pi)^D$ etc. coming from Fourier transforms
as before.\\
We are in a position, now, to put forward an argument that there is one-to-one
correspondence between the class of functions, $F_C(q)$, 
( this  is in the class ${\cal C}$) and
solution of the wave equation in the ${\tilde r}$-space,
${\bar F}({\tilde r})$ which is endowed with a rotational symmetry in
the $p_1$-$p_2$ plane. Notice the representation of ${\bar F}({\tilde r})$ is
expressed in terms of $F_C(q)$. Our desired goal is to choose a suitable
${\bar F}(u,{\bar s})$ and choose the surface, $\Sigma$ to achieve a
representation for  $F_C(q)$ with special support in the momentum space
(which is obtained from the support properties of ${\tilde F}(x)$). Our aim
is to identify the analog of coincidence region (where $F_C(q)=0$). Following
Dyson, we define a region ${\bf R}$, in the momentum space ($q$-space) which
is bounded by two spacelike surfaces $\sigma_1$ and $\sigma_2$. To be
specific choose $\bf R$ as follows:
\be
\label{dyson25}
{\bf R}:~~~~{\bar s}_1({\bf q})<q_0<{\bar s}_2({\bf q})
\ee
Inside this domain $F_C(q)=0$. Moreover, the two surfaces are chosen in such
a way that
\bea
\label{dyson26}
&&|{\bar s}_1({\bf q}) -{\bar s}_1({\bf q}')|<|{\bf q}-{\bf q}'|\nonumber\\&&
|{\bar s}_2({\bf q})- {\bar s}_2({\bf q}')|<|{\bf q}-{\bf q}'|
\eea
Here ${\bf q}$ is the $(D-1)$ component vector along spatial directions of
the D-vector $q$ and the same definition holds 
 for ${\bf q}'$. Thus we have defined two
spacelike surfaces with $q^0= {\bar s}_1({\bf q})$ and
$q^0= {\bar s}_2({\bf q})$. Now define $C_R$ to be class of functions such
that ${\tilde F}_C(x)=0$ for $x^2<0$ and such that $F_C(q)=0$ for any
$q \in {\bf R}$. Notice that the hyperboloid $(q-u)^2-{\bar s}=0$ is
$q$-space admissible. This property is valid if the upper sheet does not come
below $\sigma_2$ and the lower sheet is above $\sigma_1$. In the
$(D+2)$-dimensional space, the hyperboloid in question corresponds to points
${\tilde r}=(u_0,u_1,...u_{D-1}, p_1,p_2)$, ${\bar s}= p_1^2+p_2^2$ lying in a
certain region $\bf S$ of ${\tilde r}$-space (recall $\bf R$ is defined in
 $q$-space and $C_R$ in coordinate space).  Our intent is to derive a
representation for $F_C(q)$.
 Now for every $\tilde r$ in $\bf S$
but $q$ in the region $\bf R$, ${\bar D}({\tilde r}-{\hat q})$ vanishes. The
following expression is a prospective representation for $F_C(q)$ in $C_R$
\be
\label{dyson27}
F_C(q)=\int_{\Sigma}d\Sigma_{\alpha}\bigg[{\bar F}({\tilde r}),
{{\partial}\over{{\partial}{{\tilde r}_{\alpha}}}}[\epsilon(u_0-q_0)
\delta'\bigg((u-q)^2-{\bar s}\bigg)]\bigg]
\ee
Note that the points of $\tilde r$ are constrained to be in $\bf S$. 
Furthermore,
every point of $\tilde r$ and $\Sigma$ in $\bf S$ 
are required to belong to $C_R$
which follows from the conditions stated above. It is important to point out
that $F_C(q)$ has a representation using only the admissible hyperboloid
i.e. every $F_C(q)$ we intend to construct must have variables belonging to
the admissible hyperboloid: $(q-u)^2-{\bar s}=0$. Another constraint is that
this must not cross the surface defined by $q_0={\bar s}_1(\bf q)$ and
$q_0={\bar s}_2(\bf q)$ (see (\ref{dyson26})).\\
Let us focus attention at the upper sheet of the hyperboloid and it
corresponds to the branch
\be
\label{dyson28}
q_0=u_0+\sqrt{({\bf q}-{\bf u})^2+{\bar s}}
\ee
This will cross $\sigma_2$ if
\be
\label{dyson29}
u_0+{\sqrt{({\bf q}-{\bf u})^2+{\bar s}}}\ge{\bar s}_2({\bf q})
\ee
for ${\bf q}$ held fixed. We could rephrase the above constraint as
\be
\label{dyson30}
u_0\ge {\rm Max}~_{{\bf q}}~\{ {\bar s}_2({\bf q})-
{\sqrt{({\bf q}-{\bf u})^2+{\bar s}}}\}=m({\bf u},{\bar s})
\ee
We can repeat the same steps for the lower sheet and obtain
\be
\label{dyson31}
u_0\le {\rm Min}~_{{\bf q}}~\{ {\bar s}_1({\bf q})+
{\sqrt{({\bf q}-{\bf u})^2+{\bar s}}}\}=M({\bf u},{\bar s})
\ee
We have closely followed Dyson's notation and convention. The principal
reason is that unlike Jost-Lehmann representation which was derived for 
the case of equal
masses, the advantage of Dyson's formulation lies in the fact
 that the case of unequal mass
is treated elegantly and the approach is quite general.
 The connection with Jost-Lehmann formulation will be clear later. We have
mentioned earlier that the points on the hyperboloid correspond 
to region $\bf S$
(see the remark preceding (\ref{dyson27}) where we define region $\bf S$ in
$\tilde r$-space). For the present consideration the region $\bf S$, in the
$\tilde r$-space, can be identified to be
\be
\label{dyson32}
m({\bf u},{\bar s})\le u_0\le M({\bf u},{\bar s})
\ee
and it is bounded by two surfaces $\Sigma_1$ and  $\Sigma_2$ in the
$\tilde r$-space. We mention in passing that these surfaces are envelopes of
two families of hyperboloids and these two are also spacelike. Now define
$\bf T$: complements of $\bf S$ i.e. it contains the set of points in the
$\tilde r$-space such that
\be
\label{dyson33}
M({\bf u},{\bar s})\le u_0\le m({\bf u},{\bar s})
\ee
The purpose is to impose a constraint on ${\bar F}({\tilde r})$ in order that
representation for $F_C(q)$ gives an $F_C(q)$ such that its Fourier transform
belongs to a class which is in $C_R$. In order that  this condition is
fulfilled ${\bar F}_C({\tilde r})$ must vanish for each $\tilde r$ in $\bf T$.\\
Now, for equation (\ref{dyson20}), choose a spacelike surface, $\Sigma$,
such that it lies between the two spacelike surfaces
 $\Sigma_1$ and  $\Sigma_2$.  This surface is identified to be
\be
\label{dyson34}
u_0={1\over 2}[ m({\bf u},{\bar s})+M({\bf u},{\bar s})]
\ee
We have already constrained $u_0$ to lie in the regions given by
(\ref{dyson30}) and (\ref{dyson31}) and it also chosen to be (\ref{dyson34}).
 Therefore, every point of the chosen spacelike surface,
$\Sigma$, is either in the domain $\bf S$ or it lies in its complement $\bf T$.
According to stipulation ${\bar F}({\tilde r})$ is required to vanish for
every ${\tilde r} \in {\bf T}$. A function $F_C(q)$ belongs to $C_R$
( the Fourier transform is meant to be in $C_R$ ) if and only if it admits
a unique representation
\be
\label{dyson35}
F_C(q)= \int_{\Sigma}d\Sigma_{\alpha}\bigg[{\bar F}({\tilde r}),
{{\partial}\over{{\partial}{{\tilde r}_{\alpha}}}}D({\tilde r}-{\hat q})\bigg]
\ee
where $\Sigma\in \bf S$, in other words this integral extends only those points
of  ${\tilde r}$ of the spacelike surface  $\Sigma$ which belong to $\bf S$. We
recall that the set of points in domain $\bf S$ are given by (\ref{dyson30}) and
(\ref{dyson31}) and ${\bf S}:~ m({\bf u},{\bar s})\le q_0 \le M({\bf u},{\bar s})$.
Thus generalized version, for D-dimensional case, is\\
{\it Theorem}: For a function $F_C(q)$ to vanish in the region
${\bar s}_1({\bf q})< q_0 < {\bar s}_2({\bf q})$ and to have a Fourier
transform, ${\tilde f}(x)$ such that ${\tilde f}(x)=0$ for $x^2<0$, it is
necessary and sufficient to have a representation
\be
\label{dyson36}
F_C(q)=\int d^Du\int_0^{\infty}\epsilon(q_0-u_0)
\delta[({\bf q}-{\bf u})^2-{\bar s}]{\bf\Phi}(u,{\bar s})
\ee
${\bf\Phi}(u,{\bar s})$ vanishes outside the regions (
$u_0\ge {\rm Max}~_{{\bf q}}~\{ {\bar s}_2({\bf q})-
{\sqrt{({\bf q}-{\bf u})^2+{\bar s}}}\}$ and
$u_0\le {\rm Min}~_{{\bf q}}~\{ {\bar s}_1({\bf q})+
{\sqrt{({\bf q}-{\bf u})^2+{\bar s}}}\}$ and already noted earlier)  and
${\bf S}:~m({\bf u},{\bar s})\le u_0\le M({\bf u},{\bar s})$, but arbitrary
otherwise. Note that ${\bf\Phi}(u,{\bar s})$, appearing in (\ref{dyson36}),
depends on $q$'s determined by $(u-q)^2={\bar s}$ which lie entirely in
${\bf R}$. It reproduces the function, $F_C(q)$ on the left hand side of
(\ref{dyson36}) with the requisite support properties in $q$-space and the
support properties of ${\tilde F}_C(x)$ are satisfied. Thus we can write
\be
\label{dyson37}
{\tilde F}_C(x)= \int_0^{\infty}d{\bar s}\Delta(x;{\bar s})
{\bf\Phi}(x,{\bar s})
\ee
where ${\bf\Phi}(x,{\bar s})$ is the Fourier transform of
${\bf\Phi}(u,{\bar s})$ with respect to $u$ and is the well known invariant
function (now defined in D-dimensions) with {\it mass} ${\sqrt {\bar s}}$.
Thus the causality properties of ${\tilde F}_C(x)$, as desired by us,
is satisfied.\\
Let us consider a specific situation to make connections with our derivation
of the Jost-Lehmann representation in Section 3. Following Dyson, we choose
the two surfaces $\sigma_1$ and $\sigma_2$ to be
\bea
\label{dyson38}
&&{\bar s}_1({\bf q})=a-{\sqrt{{\bf q}^2+m_2^2}}\nonumber\\&&
{\bar s}_2({\bf q})=-a+{\sqrt{{\bf q}^2+m_1^2}}
\eea
We keep the masses $m_1$ and  $m_2$ unequal. The region $\bf S$ is identified
to be
\bea
\label{dyson39}
&& m({\bf u},{\bar s})={\rm Max}~_{\bf q}~\{ \sqrt{{\bf q}^2+m_1^2}\},
-a-{\sqrt{({\bf q}-{\bf u})^2+{\bar s}}}\nonumber\\&&
M({\bf u},{\bar s})={\rm Max}~_{\bf q}~\{ +\sqrt{{\bf q}^2+m_2^2}\},
-a-{\sqrt{({\bf q}-{\bf u})^2+{\bar s}}}\
\eea
The extremum of $m({\bf u},{\bar s})$ and $M({\bf u},{\bar s})$ is
derived by taking their gradients with respect to $\bf q$ and set each
of the gradient to zero and derive the locations of maxima.
\\
In order to establish connections with the Jost-Lehmann representation,
we identify (recall ${1\over 2}(Q_i+Q_f))$) and $m_1=m_2=m$ then domain
$\bf S$ is $(Q+q)\in V^+$, $(Q-q)\in V^+$, ${\bar s}=\chi^2$. Note,
${\bar s}=p_1^2+p_2^2$, defined in terms of the momenta along extra directions
and there is integration over $d{\bar s}$ in the expression for $F_C(q)$,
(\ref{dyson37}). Indeed  in the case of equal mass scattering we get back the
result derived by the techniques of Jost and Lehmann. We know how to derive
the presentation for the retarded function $F_R(q)$ since the two are
simply related in their coordinate space definition:
${\bar F}_C(x)= \theta (x_0){\tilde F}_C(x)$. The power of the 
mathematical approach
of Dyson is quite evident and its generalization to D-dimensions
is achieved in a very elegant manner.

\bigskip

\noindent{\bf Appendix B: Martin's Lemma}
                           
\bigskip
\noindent In this appendix we prove two lemmas which are very useful to
prove Martin's theorem. We mention that, essentially they deal with interchange
of differentiation and integration in order to prove certain analyticity properties of
functions which depend on two variables. Therefore, the proof of these lemmas
are not dependent on the dimensionality of spacetime. As will be obvious,
these functions, in nutshell, depend on two variables, and these
are to be identified with the Mandelstam variables $s$ and $t$ eventually
when the Martin's theorem is discussed.\\
Lemma 1. Suppose $F(t)=\int_0^{\infty}dsG(s,t)$ and it fulfills following
requirements:\\
(a) $F(t)$ is analytic in the neighborhood of $-\alpha\le t\le 0$.\\
(b) For all $s\in[\beta,\infty)$, $G(s,t)$ in analytic in $t$ in the
neighborhood of $-\alpha \le t\le 0$. The $t$-derivatives of $G(s,t)$ satisfy
 the condition that $({{\partial}\over{\partial t}})^nG(s,t)$ are bounded by
some functions $G_n(s)$.\\
(c) For all
$s\in [\beta,\infty):|({{\partial}\over{\partial t}})^nG(s,t)|\le 
({{\partial}\over{\partial t}})^nG(s,t)|_{t=0}$ in the interval
$-\alpha\le t\le 0$\\
Then
\be
\label{1lemma1}
({{\partial}\over{\partial t}})^nF(t)=\int_{\beta}^{\infty}
({{\partial}\over{\partial t}})^nG(s,t)ds
\ee
for $n=0,1,2...$ and for $t$ in the interval $-\alpha\le t\le 0$. What is the
purpose of this lemma? It is to prove, under what conditions, the operations
of differentiation in variable $t$ and the integration in variable $s$
 can the interchanged as we shall see.\\
Proof:  In the interval for $-\alpha\le t\le 0$, assumption (a) implies that
\bea
\label{1lemma2}
{{{\partial}\over{\partial t}}}F(t)=&&{{{\partial}\over{\partial t}}}
\int_{\beta}^{\infty}dsG(s,t)\nonumber\\&&
={\rm lim}_{\epsilon\rightarrow 0}\int_{\beta}^{\infty}ds{{G(s,t)-G(s,t-\epsilon})
\over{\epsilon}}
\eea
exists (moreover, it is analytic in the closed interval $-\alpha\le t\le 0$).
Now we appeal to the assumption (b) regarding the analyticity property of
$G(s,t)$. We are required to show that the limit $\epsilon\rightarrow 0$
and the integral can be interchanged. In order to accomplish this goal, we
write
\bea
\label{1lemma3}
{{\partial}\over{\partial t}}F(t)=\int_{\beta}^{S_1}ds
{{\partial}\over{\partial t}}G(s,t)+{\rm lim}_{\epsilon\rightarrow 0}
\int_{S_1}^{\infty}{{G(s,t)-G(s,t-\epsilon})\over{\epsilon}}
\eea
We are permitted to carry out this operation due to  the following
reasons:  the integral has a limit and it is bounded by finite function of
$s$. Therefore, the (Lebesgue) integral over a finite interval converges.
Thus, what remains is to be demonstrated is  that for the second integral of
(\ref{1lemma3}). i.e $\int_{S_1}^{\infty}$, the limit and integration can be
interchanged. We argue, invoking (c), to achieve this
\bea
\label{1lemma4}
0\le &&\bigg|\int_{S_1}^{\infty}{{G(s,t)-G(s,t-\epsilon)}\over{\epsilon}}\bigg|
\nonumber\\&&
=\bigg|\int_{S_1}^{\infty}{{\partial}\over{\partial t}}G(s,t-\epsilon')\bigg|
\le \int_{S_1}^{\infty}\bigg|({{{\partial}\over{\partial t}}})G(s,t-\epsilon')
\bigg|\nonumber\\&&
\le \int_{S_1}^{\infty} {{\partial}\over{\partial t}} G(s,t=0)
\eea
When we write the last term, it is already understood that the limit
$\epsilon\rightarrow 0$ has been taken at the appropriate stage.
\\
Lemma 2: Consider two functions $G_1(s,t)$ and $G_2(s,t)$ which fulfill the
requirements (b) and (c) of Lemma 1 then the product also have the same
properties.\\
Proof: Define
\be
\label{2lemma1}
G(t)=\int_{\beta}^{\infty} dsG_1(s,t)G_2(s,t)
\ee
and it is analytic in $|t|<R$. Then for all complex $z\notin [\beta,\infty)$
and it is analytic in $|t|<R$. Consequently, for all complex $z\notin [\beta,\infty)$
\bea
\label{2lemma2}
\bigg|{{ ({{\partial}\over{\partial t}})^nG_1(s,0)}\over{s-z}}\bigg|\le&&
{\rm Sup}_{\beta\le s\le\infty}~\bigg|{{1}\over{(s-z)G_2(s,0)}}\bigg|
\int_{\beta}^{\infty}dsG_2(s,0)\bigg[{ ({{\partial}\over{\partial t}})^n
G_1(s,t)}\bigg]_{t=0}\nonumber\\&&
\le{\rm Sup}_{\beta\le s\le\infty}~\bigg|{1\over{(s-z)G_2(s,0)}}\bigg|
\int_{\beta}^{\infty}\bigg[ ({{\partial}\over{\partial t}})^n
\bigg(G_1(s,t)G_2(s,t)\bigg)
\bigg]_{t=0}\nonumber\\&&
\le {{n!}\over{R^n}}~{\rm Sup}_{\beta\le s\le\infty}~\bigg|{1\over{(s-z)G_2(s,0)}}
\bigg|~{\rm Max}_{|t|<R}~G(t)
\eea
The first inequality obviously follows from the properties of function. The
next one is a consequence of the requirement (c) stated in Lemma 1. The last
inequality is due to application of the Cauchy's inequality for the function
$G(t)$ defined above.\\

\bigskip

\noindent{\bf Appendix C: Useful formulas used for the Gegenbauer Polynomial}

\bigskip
\noindent  We compile some of the useful formulas used in this article. We
give volume and page number of the Batesman manuscript - the exact reference
is given in the reference section. In our case $\lambda={1\over 2}(D-3)$
where D is number of spacetime dimensions. We are dealing with higher
spacetime dimensions i.e. $D>$. Therefore, $\lambda \ge 1$\\
Orthogonal polynomials.\\
\be
\label{op1}
 (\phi_(x),\phi_2(x))=\int_a^b{\cal W}(x)\phi_1(x)\phi_2(x)dx
\ee
For Gegebauer polynomials $C^{\lambda}_l$, ~~$a=-1$,~~ $b=+1$,~~
${\cal W}(x) =(1-x^2)^{\lambda-1/2} $
\\
The formulas from Vol I. \\
p175 
\be
C^{\lambda}_n(z)=\sum_{l=0}^n{{(-1)^l\Gamma(\lambda+l)\Gamma(n+2\lambda+l)}
\over{l!(n-l)!\Gamma(\lambda)\Gamma(2l+\lambda}}({1\over 2}-{1\over 2}z)^l
\ee
p176 
\be
{({d\over{dz}})^n}[C^{\lambda}_n(z)]=
2^n{{\Gamma(\lambda+n)}\over{\Gamma(\lambda)}}
\ee
p176 
\be
{1\over 2}(\Gamma(\lambda))^2C^{\lambda}_n(cos\phi)\sum_{m=0}^{\le n/2}
{{\Gamma(m+\lambda)\Gamma(n-m+\lambda)cos[(n-2m)\phi]}\over{m!(n-m)!}}
\ee
\be
C^{\lambda}_n(x)={1\over{\sqrt\pi}}{{\Gamma(n+2\lambda)\Gamma(\lambda+1/2)}
\over{{\Gamma(\lambda)\Gamma(2\lambda)\Gamma(n+1)}}}
\int_0^{\pi}\bigg[x+\sqrt{(x^2-1)}cos\phi \bigg]^{\lambda}
(sin\phi)^{2\lambda-1}d\phi
\ee

p178 
\be
{{d\over{dz}}}C^{\lambda}_n(z)=2\lambda C^{\lambda+1}_{n-1}(z)
\ee

\noindent
Vol II.\\
Inequality: p206
\be
{\rm Max}_{-1\le z\le +1}|C^{\lambda}_n(z)|=C^{\lambda}_n(1)>0
\ee

\end{document}